\documentclass[10pt,draftcls,onecolumn]{IEEEtran} 
\usepackage{epsfig,algorithm,amsmath,amssymb,color,subfigure,url}
\author{
\authorblockN{Namrata Vaswani
}
\authorblockA{ECE Dept., Iowa State University, Ames, IA 50011, USA, Email: namrata@iastate.edu
}
\thanks{This work was supported by NSF grants ECCS-0725849, CCF-0917015.}
}

\title{Stability (over time) of Modified-CS and LS-CS for Recursive Causal Sparse Reconstruction} %: Part 2

\begin{document}
\setlength{\arraycolsep}{0.03cm}
\newcommand{\xhat}{\hat{x}}
\newcommand{\xpred}{\hat{x}_{t|t-1}}
\newcommand{\Ppred}{P_{t|t-1}}
\newcommand{\ty}{\tilde{y}_t}
\newcommand{\tty}{\tilde{y}_{t,\text{res}}}
\newcommand{\tw}{\tilde{w}_t}
\newcommand{\ttw}{\tilde{w}_{t,f}}
\newcommand{\betahat}{\hat{\beta}}

\newcommand{\ypast}{y_{1:t-1}}
\newcommand{\sone}{S_{*}}
\newcommand{\sinf}{{S_{**}}}
\newcommand{\smax}{S_{\max}}
\newcommand{\smin}{S_{\min}}
\newcommand{\samax}{S_{a,\max}}
\newcommand{\Nhat}{{\hat{N}}}

\newcommand{\sgn}{\text{sgn}}

\newcommand{\Dnum}{D_{num}}
\newcommand{\pss}{p^{**,i}}
\newcommand{\fr}{f_{r}^i}

\newcommand{\A}{{\cal A}}
\newcommand{\Z}{{\cal Z}}
\newcommand{\B}{{\cal B}}
\newcommand{\R}{{\cal R}}
\newcommand{\reg}{{\cal G}}
\newcommand{\const}{\mbox{const}}

\newcommand{\trace}{\mbox{tr}}

\newcommand{\hsim}{{\hspace{0.0cm} \sim  \hspace{0.0cm}}}
\newcommand{\he}{{\hspace{0.0cm} =  \hspace{0.0cm}}}

\newcommand{\vect}[2]{\left[\begin{array}{cccccc}
     #1 \\
     #2
   \end{array}
  \right]
  }

\newcommand{\matr}[2]{ \left[\begin{array}{cc}
     #1 \\
     #2
   \end{array}
  \right]
  }
\newcommand{\vc}[2]{\left[\begin{array}{c}
     #1 \\
     #2
   \end{array}
  \right]
  }

\newcommand{\gdot}{\dot{g}}
\newcommand{\Cdot}{\dot{C}}
\newcommand{\re}{\mathbb{R}}
\newcommand{\n}{{\cal N}}  %normal distribution
\newcommand{\N}{{\overrightarrow{\bf N}}}  % normal to contour
\newcommand{\chat}{\tilde{C}_t}
\newcommand{\chati}{\chat^i}

\newcommand{\cmin}{C^*_{min}}
\newcommand{\twi}{\tilde{w}_t^{(i)}}
\newcommand{\twj}{\tilde{w}_t^{(j)}}
\newcommand{\wi}{{w}_t^{(i)}}
\newcommand{\twio}{\tilde{w}_{t-1}^{(i)}}

\newcommand{\tWi}{\tilde{W}_n^{(m)}}
\newcommand{\tWj}{\tilde{W}_n^{(k)}}
\newcommand{\Wi}{{W}_n^{(m)}}
\newcommand{\tWio}{\tilde{W}_{n-1}^{(m)}}

\newcommand{\ds}{\displaystyle}

\newcommand{\SAR}{S$\!$A$\!$R }
\newcommand{\MAR}{MAR}
\newcommand{\MMRF}{MMRF}
\newcommand{\AR}{A$\!$R }
\newcommand{\GMRF}{G$\!$M$\!$R$\!$F }
\newcommand{\DTM}{D$\!$T$\!$M }
\newcommand{\MSE}{M$\!$S$\!$E }
\newcommand{\RCS}{R$\!$C$\!$S }
\newcommand{\uomega}{\underline{\omega}}
\newcommand{\y}{v}
\newcommand{\x}{w}
\newcommand{\lu}{\mu}
\newcommand{\g}{g}
\newcommand{\bft}{{\bf t}}
\newcommand{\refmap}{{\cal R}}
\newcommand{\totrefl}{{\cal E}}
\newcommand{\beq}{\begin{equation}}
\newcommand{\eeq}{\end{equation}}
\newcommand{\bdm}{\begin{displaymath}}
\newcommand{\edm}{\end{displaymath}}
\newcommand{\hatz}{\hat{z}}
\newcommand{\hatu}{\hat{u}}
\newcommand{\tilz}{\tilde{z}}
\newcommand{\tilu}{\tilde{u}}
\newcommand{\hhatz}{\hat{\hat{z}}}
\newcommand{\hhatu}{\hat{\hat{u}}}
\newcommand{\tilc}{\tilde{C}}
\newcommand{\hatc}{\hat{C}}
\newcommand{\tim}{n}

\newcommand{\ssp}{\renewcommand{\baselinestretch}{1.0}}
\newcommand{\defd}{\mbox{$\stackrel{\mbox{$\triangle$}}{=}$}}
\newcommand{\goes}{\rightarrow}
\newcommand{\tends}{\rightarrow}
\newcommand{\defn}{\triangleq} %{\stackrel{\triangle}{=}}
\newcommand{\se}{&=&}
\newcommand{\sdefn}{& \defn  &}
\newcommand{\sle}{& \le &}
\newcommand{\sge}{& \ge &}
\newcommand{\plusminus}{\stackrel{+}{-}}
\newcommand{\Ey}{E_{Y_{1:t}}}
\newcommand{\ey}{E_{Y_{1:t}}}

\newcommand{\equivto}{\mbox{~~~which is equivalent to~~~}}
\newcommand{\nonzero}{i:\pi^n(x^{(i)})>0}
\newcommand{\nonzeroc}{i:c(x^{(i)})>0}

\newcommand{\supn}{\sup_{\phi:\|\phi\|_\infty \le 1}}

\newtheorem{theorem}{Theorem}
\newtheorem{lemma}{Lemma}
\newtheorem{corollary}{Corollary}
\newtheorem{definition}{Definition}
\newtheorem{remark}{Remark}
\newtheorem{example}{Example}
\newtheorem{ass}{Assumption}
\newtheorem{proposition}{Proposition}

\newtheorem{fact}{Fact}
\newtheorem{heuristic}{Heuristic}
\newcommand{\eps}{\epsilon}
\newcommand{\bd}{\begin{definition}}
\newcommand{\ed}{\end{definition}}
\newcommand{\udq}{\underline{D_Q}}
\newcommand{\td}{\tilde{D}}
\newcommand{\epsinv}{\epsilon_{inv}}
\newcommand{\al}{\mathcal{A}}

\newcommand{\bfx} {\bf X}
\newcommand{\bfy} {\bf Y}
\newcommand{\bfz} {\bf Z}
\newcommand{\ddas}{\mbox{${d_1}^2({\bf X})$}}
\newcommand{\ddbs}{\mbox{${d_2}^2({\bfx})$}}
\newcommand{\dda}{\mbox{$d_1(\bfx)$}}
\newcommand{\ddb}{\mbox{$d_2(\bfx)$}}
\newcommand{\xinc}{{\bfx} \in \mbox{$C_1$}}
\newcommand{\eqa}{\stackrel{(a)}{=}}
\newcommand{\eqb}{\stackrel{(b)}{=}}
\newcommand{\eqe}{\stackrel{(e)}{=}}
\newcommand{\leqc}{\stackrel{(c)}{\le}}
\newcommand{\leqd}{\stackrel{(d)}{\le}}

\newcommand{\leqa}{\stackrel{(a)}{\le}}
\newcommand{\leqb}{\stackrel{(b)}{\le}}
\newcommand{\leqe}{\stackrel{(e)}{\le}}
\newcommand{\leqf}{\stackrel{(f)}{\le}}
\newcommand{\leqg}{\stackrel{(g)}{\le}}
\newcommand{\leqh}{\stackrel{(h)}{\le}}
\newcommand{\leqi}{\stackrel{(i)}{\le}}
\newcommand{\leqj}{\stackrel{(j)}{\le}}

\newcommand{\w}{{W^{LDA}}}
\newcommand{\halpha}{\hat{\alpha}}
\newcommand{\hsigma}{\hat{\sigma}}
\newcommand{\slmax}{\sqrt{\lambda_{max}}}
\newcommand{\slmin}{\sqrt{\lambda_{min}}}
\newcommand{\lmax}{\lambda_{max}}
\newcommand{\lmin}{\lambda_{min}}

\newcommand{\da} {\frac{\alpha}{\sigma}}
\newcommand{\chka} {\frac{\check{\alpha}}{\check{\sigma}}}
\newcommand{\sumo}{\sum _{\underline{\omega} \in \Omega}}
\newcommand{\distance}{d\{(\hatz _x, \hatz _y),(\tilz _x, \tilz _y)\}}
\newcommand{\col}{{\rm col}}
\newcommand{\rcs}{\sigma_0}
\newcommand{\CalR}{{\cal R}}
\newcommand{\df}{{\delta p}}
\newcommand{\dq}{{\delta q}}
\newcommand{\dZ}{{\delta Z}}
\newcommand{\pprime}{{\prime\prime}}

\newcommand{\vn}{N}

\newcommand{\bv}{\begin{vugraph}}
\newcommand{\ev}{\end{vugraph}}
\newcommand{\bi}{\begin{itemize}}
\newcommand{\ei}{\end{itemize}}
\newcommand{\ben}{\begin{enumerate}}
\newcommand{\een}{\end{enumerate}}
\newcommand{\be}{\protect\[}
\newcommand{\ee}{\protect\]}
\newcommand{\bean}{\begin{eqnarray*} }
\newcommand{\eean}{\end{eqnarray*} }
\newcommand{\bea}{\begin{eqnarray} }
\newcommand{\eea}{\end{eqnarray} }
\newcommand{\nn}{\nonumber}
\newcommand{\ba}{\begin{array} }
\newcommand{\ea}{\end{array} }
\newcommand{\ep}{\mbox{\boldmath $\epsilon$}}
\newcommand{\epp}{\mbox{\boldmath $\epsilon '$}}
\newcommand{\Lep}{\mbox{\LARGE $\epsilon_2$}}
\newcommand{\und}{\underline}
\newcommand{\pdif}[2]{\frac{\partial #1}{\partial #2}}
\newcommand{\odif}[2]{\frac{d #1}{d #2}}
\newcommand{\dt}[1]{\pdif{#1}{t}}
\newcommand{\urho}{\underline{\rho}}

\newcommand{\spc}{{\cal S}}
\newcommand{\tspc}{{\cal TS}}

\newcommand{\uv}{\underline{v}}
\newcommand{\us}{\underline{s}}
\newcommand{\uc}{\underline{c}}
\newcommand{\utheta}{\underline{\theta}^*}
\newcommand{\ualpha}{\underline{\alpha^*}}

\newcommand{\uxy}{\underline{x}^*}
\newcommand{\uxyj}{[x^{*}_j,y^{*}_j]}
\newcommand{\arcl}[1]{arclen(#1)}
\newcommand{\one}{{\mathbf{1}}}

\newcommand{\uxyjt}{\uxy_{j,t}}
\newcommand{\E}{\mathbb{E}}

\newcommand{\rhomat}{\left[\begin{array}{c}
                        \rho_3 \ \rho_4 \\
                        \rho_5 \ \rho_6
                        \end{array}
                   \right]}
\newcommand{\deltat}{\tau} %{\Delta t}
\newcommand{\deltatt}{\Delta t_1}
\newcommand{\ceil}[1]{\ulcorner #1 \urcorner}

\newcommand{\xxi}{x^{(i)}}
\newcommand{\txi}{\tilde{x}^{(i)}}
\newcommand{\txj}{\tilde{x}^{(j)}}

\newcommand{\mi}[1]{{#1}^{(m,i)}}

\setlength{\arraycolsep}{0.05cm}
\newcommand{\rest}{{T_\text{rest}}}
\newcommand{\zetahat}{\hat{\zeta}}
\newcommand{\tDelta}{{\tilde{\Delta}}}
\newcommand{\tDeltae}{{\tilde{\Delta}_e}}
\newcommand{\tT}{{\tilde{T}}}
\newcommand{\add}{{\cal A}}
\newcommand{\rem}{{\cal R}}
\newtheorem{sigmodel}{Signal Model}

\newcommand{\thr}{{\text{thr}}}
\newcommand{\delthr}{{\text{del-thr}}}
\newcommand{\delbound}{{b}}
\newcommand{\err}{{\text{err}}}
\newcommand{\Q}{{\cal Q}}

\newcommand{\dett}{{\text{add}}}  %{{\text{det}}}
\newcommand{\CSres}{{\text{CSres}}}
\newcommand{\diff}{{\text{diff}}}
\newcommand{\Section}[1]{ \vspace{-0.13in}  \section{#1} \vspace{-0.12in} } %  \vspace{-0.1in}   %\vspace{-0.05in}
\newcommand{\Subsection}[1]{  \vspace{-0.12in} \subsection{#1}  \vspace{-0.08in} } %  \vspace{-0.1in}   %\vspace{-0.05in}
\newcommand{\Subsubsection}[1]{   \subsubsection{#1} } %  \vspace{-0.1in}   %\vspace{-0.05in}

\date{}
\maketitle

\newcommand{\Aset}{{\cal A}}
\newcommand{\Rset}{{\cal R}}
\newcommand{\Iset}{{\cal I}}
\newcommand{\Dset}{{\cal D}}
\newcommand{\Sset}{{\cal S}}
\newcommand{\Inc}{\text{Inc}}
\newcommand{\Dec}{\text{Dec}}
\newcommand{\Con}{\text{Con}}
\newcommand{\sm}{e}  %{\|x_{\Sset}\|^2}

\begin{abstract}
%In this work, we show the ``stability" of two of our recently proposed algorithms, LS-CS-residual (LS-CS) and the noisy version of modified-CS, designed for recursive reconstruction of sparse signal sequences from noisy measurements.
In this work, we obtain sufficient conditions for the ``stability" of our recently proposed algorithms,  modified-CS (for noisy measurements) and Least Squares CS-residual (LS-CS), designed for recursive reconstruction of sparse signal sequences from noisy measurements. By ``stability" we mean that the number of misses from the current support estimate and the number of extras in it remain bounded by a time-invariant value at all times. The concept is meaningful only if the bound is small compared to the current signal support size. A direct corollary is that the reconstruction errors are also bounded by a time-invariant and small value.%
%
% We show that for both our algorithms, ``stability" holds under mild assumptions (bounded noise, high enough SNR and enough measurements at every time) for a simple deterministic signal model with fixed signal power and support size, slow support changes occurring at every time, and gradual coefficient magnitude increase/decrease. Under certain situations, the result for modified-CS is stronger. A direct corollary is that the reconstruction errors are also bounded by a constant at all times (``stable").
% (compressive sensing on least squares residual)
% remains bounded by a small value compared to the current signal support size. This, in turn, implies that the LS-CS reconstruction error remains bounded by a small value compared to the total signal power.
%From Theorem \ref{stabres2}, under Signal Model \ref{sigmod2}, ``stability" is ensured under the following mild assumptions: (a) the addition/deletion thresholds are appropriately set (conditions \ref{addthresh} and \ref{delthresh});  (b) the noise is bounded and (c) the SNR is high enough and there are enough measurements (for a given support change rate, $S_a$, and magnitude change rate, $M/d$) so that $\sone, \sinf$ are large enough for condition \ref{measmodel} to hold and $\theta^2$, $C',C''$ are small enough for condition \ref{add_del} to hold.%By stability, we mean that $|\tDelta|$, $|\tDelta_e|$ remain bounded by small enough values compared to the signal support size.
%(if $d_0 S_a$ and $f$ are small compared to $S_0$). $|\tDelta_t|$ and $|\tDelta_{e,t}|$ are
\end{abstract}

\section{Introduction} % {\em in real-time}  such as the beating heart  compressive sensing on least squares residual
In this work, we study the ``stability" of modified-CS (noisy) \cite{modcsjp,modcsicassp10} and of LS-CS-residual (LS-CS) \cite{just_lscs,kfcsicip,kfcspap} which were designed for recursive reconstruction of sparse signal sequences from noisy measurements. By ``stability" we mean that the number of misses from the current support estimate and the number of extras in it remain bounded by a {\em time-invariant} value at all times. The concept is meaningful only if the bound is small compared to the current signal support size. A direct corollary is that the reconstruction errors are also bounded by a time-invariant and small value.%

The key assumption that our algorithms utilize is that the support changes slowly over time. As we demonstrated in \cite{just_lscs,modcsjp}, this assumption holds for many medical image sequences. Denote the support estimate from the previous time by $T$. Modified-CS tries to finds a signal that is sparsest outside of $T$ and satisfies the data constraint. LS-CS uses a different approach. It replaces compressive sensing (CS) on the observation by CS on the least squares (LS) residual computed using  $T$. Both algorithms are able to achieve greatly reduced reconstruction error compared to simple CS (CS at each time separately) when using fewer measurements than what CS needs.%for exact reconstruction.modified-CS and LS-CS
%In ongoing work, we have demonstrated that modified-CS outperforms LS-CS when the size of the unknown support is larger, while the opposite is true when the number of extras in $T$ and the noise are larger. But both significantly outperform simple CS with fewer measurements.

%\cite{reddy} (only reconstructs the difference images from the observation differences, but can also be adapted to also recursively reconstruct the actual signal using fewer measurements, but it is not stable even in experiments);

%on Least Squares CS-residual (LS-CS) and
Other algorithms for recursive reconstruction include our older work on Kalman filtered CS-residual (KF-CS) \cite{kfcsicip,kfcspap}; CS for time-varying signals \cite{giannakis_2} (assumes a time-invariant support, which is a somewhat restrictive assumption); homotopy methods \cite{romberg} (use past reconstructions to speed up current optimization but not to improve reconstruction error with fewer measurements); and \cite{ibm} (a recent modification of KF-CS). Two other algorithms that are also designed for static CS with partial knowledge of support include \cite{camsap07} and \cite{hassibi}. The work of \cite{camsap07} proposed an approach similar to modified-CS but did not analyze it and also did not show real experiments either. The work of \cite{hassibi}, which appeared in parallel with modified-CS, assumed a probabilistic prior on the support and obtained conditions for exact reconstruction.% (proposed earlier, but we recently learnt about it)

To the best of our knowledge, stability of recursive sparse reconstruction algorithms has not been studied in any other work except in our older works \cite{just_lscs,kfcspap} for LS-CS and KF-CS respectively. %The KF-CS result was under very strong assumptions, e.g. it only handled support additions, not removals.
%The only other stability results for recursive causal sparse reconstruction that we are aware of are from our older work - LS-CS \cite{just_lscs} and KF-CS \cite{kfcspap}.
The KF-CS result \cite{kfcspap} is under fairly strong assumptions, e.g. it is for a random walk signal change model with only support additions (no removals). The result for LS-CS stability \cite{just_lscs} holds under mild assumptions and is for a fairly realistic signal change model. The only limitation is that it assumes that support changes occur ``every-so-often": every $d$ time units, there are $S_a$ support additions and $S_a$ removals. %Stability holds when $k$ is larger than a certain threshold which depends on $S_a$ and other things.
 But from testing the slow support change assumption for real data (medical image sequences), it has been observed that support changes usually occur at {\em every} time, e.g. see Fig. 1 of \cite{just_lscs}. {\em This important case is the focus of the current work.} %The overall approach that we use is motivated by that of \cite{just_lscs}.

In \cite{just_lscs}, we only studied LS-CS (modified-CS was proposed later). But the techniques of \cite{just_lscs} can be also used to show modified-CS stability for the model of \cite{just_lscs}.
In this work, we show the stability of both LS-CS and modified-CS and of its improved version, ``modified-CS with add-LS-del". We first discuss modified-CS since, from experiments, it is known to be a better algorithm. In facts its stability result is also better (holds under weaker assumptions).
% first study modified-CS stability in detail since our earlier experiments indicate that it is a better algorithm \cite{modcsjp}. LS-CS stability is also shown.
%But, again, the same techniques can be used to show LS-CS stability as well for the current signal model. This is done in the long version of this work \cite{long}.% we get a much better stability result for it and
 %Also, modified-CS stability was not studied in \cite{just_lscs}.% We are in fact able to prove a stronger result for modified-CS than that for LS-CS.%We are able to prove a stronger result for modified-CS than for LS-CS.
%(one with bounded signal power and support size and which allows support additions and removals)  and then it shows stability under the assumption that $k$ is large enough

The paper is organized as follows. We give problem definition in Sec. \ref{notn} and we overview our results in Sec. \ref{overview}. We describe the signal model for proving stability in Sec. \ref{signalmodel}. We obtain sufficient conditions for the stability of modified-CS and discuss the implications in Sec. \ref{simple_modcs}. We discuss some of its limitations and develop a simple modification that uses a better support estimation approach (modified-CS with add-LS-del). This support estimation approach is related to the one in \cite{cosamp,kfcsicip,just_lscs}. In Sec. \ref{addLSdel_modcs}, we show the stability of modified-CS with add-LS-del, which is more difficult to do. The result for LS-CS stability is obtained in Sec. \ref{addLSdel_lscs}. Numerical experiments are discussed in Sec. \ref{sims}. Conclusions are given in Sec. \ref{conclusions}.%

\subsection{Notation and Problem Definition}
\label{notn}
The set operations $\cup$, $\cap$, $\setminus$ have their usual meanings. $\emptyset$ denotes the empty set.
We use $T^c$ to denote the complement of a set $T$ w.r.t. $[1,m]:=[1,2,\dots m]$, i.e. $T^c := [1,m] \setminus T$. $|T|$ denotes the cardinality of $T$. For a vector, $v$, and a set, $T$, $v_T$ denotes the $|T|$ length sub-vector containing the elements of $v$ corresponding to the indices in the set $T$. $\| v \|_k$ denotes the $\ell_k$ norm of a vector $v$. {\em If just $\|v\|$ is used, it refers to $\|v\|_2$.} Similarly, for a matrix $M$, $\|M\|_k$ denotes its induced $k$-norm, while just $\|M\|$ refers to $\|M\|_2$. $M'$ denotes the transpose of $M$ and $M^\dag$ denotes the Moore-Penrose pseudo-inverse of $M$ (when $M$ is tall, $M^\dag:=(M'M)^{-1} M'$). For a fat matrix $A$, $A_T$ denotes the sub-matrix obtained by extracting the columns of $A$ corresponding to the indices in $T$.
The $S$-restricted isometry constant \cite{dantzig}, $\delta_S$, for an $n \times m$ matrix (with $n<m$), $A$, and the $S,S'$ restricted orthogonality constant \cite{dantzig}, $\theta_{S,S'}$, are as defined in \cite[eq 1.3]{dantzig} and \cite[eq 1.5]{dantzig} respectively.

We assume the following observation model:
\bea
y_t = A x_t + w_t,  \ \|w_t\| \le \eps   % \ \  \E[w_t]=0, \  \E[w_t w_t']= \sigma^2 I \ \ \ \
 %w_t \sim \n(0,\sigma_{obs}^2 I) %\underbrace{H \Phi}_
\label{obsmod}
\eea
where $x_t$ is an $m$ length sparse vector with support $N_t$, $y_t$ is the $n< m$ length observation vector at time $t$ and $w_t$ is observation noise with $\|w_t\| \le \eps$.  ``Support" refers to the set of indices of the nonzero elements of $x_t$.% of $x_t$

Our goal is to recursively estimate $x_t$ using $y_1, \dots y_t$. By {\em recursively}, we mean, use only $y_t$ and the estimate from $t-1$, $\xhat_{t-1}$, to compute the estimate at $t$.

As we explain in Sec. \ref{simple_modcs}, our algorithm need more measurements at the initial time, $t=0$. We use $n_0$ to denote the number of measurements used at $t=0$ and we use $A_0$ to denote the corresponding $n_0 \times m$ measurement matrix.
We use $\alpha$ to denote the support estimation threshold used by modified-CS and we use $\alpha_{add}, \alpha_{del}$ to denote the support addition and deletion thresholds used by modified-CS with add-LS-del and by LS-CS.

We use $\xhat_t$ to denote the final estimate of $x_t$ at time $t$ and $\Nhat_t$ to denote its support estimate. To keep notation simple, we avoid using the subscript $t$ wherever possible.
%We will use the following sets often.%
\bd[$T$, $\Delta$, $\Delta_e$]
We use $T  : = \Nhat_{t-1}$ to denote the support estimate from the previous time. %This serves as an initial estimate of the current support.% =T(t)
We use $\Delta := N_t \setminus T$ to denote the unknown part of the support at the current time and $\Delta_e := T \setminus N_t$ to denote the ``erroneous" part of $T$. We attach the subscript $t$ to the set, e.g. $T_t$ or $\Delta_t$, where necessary.%or $\Delta_{e,t}$
\ed
\bd[$\tT$, $\tDelta$, $\tDelta_e$]
We use $\tT := \Nhat_t$ to denote the final estimate of the current support; $\tDelta : = N_t \setminus \tT$ to denote the ``misses" in $\Nhat_t$ and $\tDelta_e := \tT \setminus N_t$ to denote the ``extras".%(extra elements that wrongly got added). (elements of the current support which did not get detected by the algorithm)
\ed
The sets $T_\dett, \Delta_\dett, \Delta_{e,\dett}$ are defined later in Sec. \ref{addLSdel_modcs}.

{\em If the sets $B,C$ are disjoint, then  we just write $D \cup B \setminus C$ instead of writing $(D \cup B) \setminus C$, e.g. $N_t = T \cup \Delta \setminus \Delta_e$.}

We refer to the left (right) hand side of an equation or inequality as LHS (RHS).

%For simplicity, we assume that the measurement matrix, $A$, has unit norm columns.
%
%The support, $N_t$, is assumed to change slowly over time, i.e. the additions, $|N_t \setminus N_{t-1}| \le S_a  \ll  |N_t|$ and the removals, $|N_{t-1} \setminus N_t| \le S_a  \ll  |N_t|$. This was verified in \cite{icipmodcs}. Also, we assume that $A$ is incoherent enough to satisfy $\delta(S) < 1/2$ (or pick any constant less than 1) for $S$ a little more than $|N_t|$. This is quantified later.%
% and its requirements on the support size are also much stronger

\vspace{-0.1in}
\subsection{Overview of Results}
\label{overview}

When measurements are noisy, the reconstruction errors of modified-CS (noisy) and of LS-CS have been bounded as a function of $|N_t|$, $|\Delta_t|$ and $|\Delta_{e,t}|$ in \cite{modcsicassp10,arxiv} and in \cite{just_lscs} respectively. The bound is small if $|\Delta_t|$ and  $|\Delta_{e,t}|$ are small enough. But smallness of the support errors, $\Delta_t$, $\Delta_{e,t}$, depends on the accuracy of the previous reconstruction. Thus it can happen that the error bound increases over time, and such a bound is of limited use for a recursive reconstruction problem. There is thus a need to obtain conditions under which one can show ``stability", i.e. obtain a time-invariant bound on the sizes of these support errors. Also, for the result to be meaningful, the support errors' bound needs to be small compared to the support size. %This would, then, directly imply a time-invariant and small bound on the reconstruction error. % (see Lemma \ref{modcsbnd} and Corollary \ref{modcs_cs_bnd})  to satisfy the required condition

%When measurements are noiseless, the stability result for modified-CS follows trivially from the exact reconstruction conditions (erc) which were derived in \cite{isitmodcs} in terms of $|\Delta|,|\Delta_e|,|N|$ \cite{isitmodcs}.  The error will be stable at zero for all times, $t$, if $n_0$ is large enough so that the erc's hold for $|\Delta| = |N_0|$, $|\Delta_e| =0$ and $|N|=|N_0|$ and if $n$ is large enough to ensure that the erc's hold for $|\Delta| = \max_{t} |N_t \setminus N_{t-1}|$, $|\Delta_e| = \max_t|N_{t-1} \setminus N_{t}|$ and $|N|=\max_t |N_t|$.

In this work, we study the stability of modified-CS for noisy measurements and its modification, modified-CS with add-LS-del, as well as of LS-CS. This is done under a bounded observation noise assumption and for a signal model with
\ben
\item support changes ($S_a$ additions and $S_a$ removals) occurring at every time, $t$,
\item magnitude of the newly added coefficients increases gradually, and similarly for decrease before removal,% is also gradual,%gradual decrease in magnitude before removal
\item support size is $S_0$ at all times and the signal power is also constant%at all times
\een
\begin{remark}
The reason we need the bounded noise assumption is as follows. When the noise is unbounded, e.g. Gaussian, all error bounds for CS and, similarly, all error bounds for LS-CS or modified-CS hold with ``large probability" \cite{dantzig,candes_rip,tropp,just_lscs,modcsicassp10,arxiv}. To show stability, we need the error bound for LS-CS or modified-CS to hold at all times, $0 \le t < \infty$ (this, in turn, is used to ensure that the support gets estimated with bounded error at all times). Clearly this is a zero probability event.%But any ``large probability" result will hold for all times, $0 < t < \infty$, with probability zero.% This is the reason why we cannot show stability without assuming bounded noise.%
\end{remark}

Our results have the following form. For a given number and type of measurements (i.e. for a given measurement matrix, $A$), and for a given noise bound, $\eps$, if,
\ben
\item the support estimation threshold(s) is/are large enough,
\item the support size, $S_0$, and support change size, $S_a$ are small enough,
\item the newly added coefficients increase (existing large coefficients decrease) at least at a certain rate, $r$, and
\item the initial number of measurements, $n_0$, is large enough for accurate initial reconstruction using simple CS,
\een
then the support errors are bounded by time-invariant values. In particular, we show that $|N_t \setminus \Nhat_t| \le 2S_a$ and $|\Nhat_t \setminus N_t| = 0$. Consequently the reconstruction error is also bounded by a small and time-invariant value.

A key assumption used in designing both modified-CS and LS-CS is that the signal support changes slowly over time. As shown in \cite{just_lscs,modcsjp}, this holds for real medical image sequences. For our model, this translates to $S_a \ll S_0$.

Under the slow support change assumption, clearly, $2S_a \ll S_0$, and so the support error bounds are small compared to the support size, $S_0$, making our stability results meaningful. We also compare the conditions on $S_0$ required by our results with those required by the corresponding simple CS error bounds (since simple CS is not a recursive approach these also serves as a stability result for simple CS) and argue that our results hold under weaker assumptions (allow larger values of $S_0$). The results for modified-CS, modified-CS (with add-LS-del) and LS-CS are also compared.%

\section{Signal model for studying stability}
\label{signalmodel}
The proposed algorithms {\em do not} assume any signal model. But to prove their stability, we need certain assumptions on the signal change over time. These are summarized here.

%We assume that the signal model (a) allows equal and nonzero number of additions/removals from the support at every time, (b) allows a new coefficient magnitude to gradually increase from zero, at a rate $M/d$, for a duration, $d$, and finally reach a constant value, $M$, (c) allows coefficients to gradually decrease and become zero (get removed from support) at the same rate, and (d) signals have constant signal power and support size at all times.

\begin{sigmodel} Assume the following.
\ben
% (the increasing subset of)

\item (addition) At each $t>0$, $S_a$ new coefficients get added to the support at magnitude $r$. Denote this set by $\Aset_t$.

\item (increase) At each $t>0$, the magnitude of $S_a$ coefficients which had magnitude $(j-1)r$ at $t-1$ increases to $jr$. This occurs for all $2 \le j \le d$. Thus the maximum magnitude reached by any coefficient is $M:=dr$.

\item (decrease) At each $t>0$, the magnitude of $S_a$ coefficients which had magnitude $(j+1)r$ at $t-1$ decreases to $jr$. This occurs for all $1 \le j \le (d-1)$.

\item (removal) At each $t>0$, $S_a$ coefficients which had magnitude $r$ at $t-1$ get removed from the support (magnitude becomes zero). Denote this set by $\Rset_t$.

\item (initial time) At $t=0$, the support size is $S_0$ and it contains $2S_a$ elements each with magnitude $r,2r, \dots (d-1)r$, and $(S_0-(2d-2)S_a)$ elements with magnitude $M$.%
%$S_a$ coefficients out of the support set with magnitude $M$ start decreasing at rate $M/d$ until they become zero.% (get removed from the support) at $t+d-1$.
\een
\label{sigmod2}
\end{sigmodel}

Notice that, in the above model, the size and composition of the support at any $t$ is the same as that at $t=0$.  Also, at each $t$, there are $S_a$ new additions and $S_a$ removals. The new coefficient magnitudes increase gradually at rate $r$ and do not increase beyond a maximum value $M:=dr$. Similarly for decrease. The support size is always $S_0$ and the signal power is always $(S_0-(2d-2)S_a)M^2 + 2S_a\sum_{j=1}^{d-1} j^2 r^2$.% at all $t$.%The size is $|N_t|=S_0$ and $N_t$ consists of $(d-1)S_a$ increasing coefficients and $(d-1)S_a$ decreasing coefficients, and $S_0- (2d-2)S_a$ constant coefficients with magnitude $M$.
%Also, at all  $t$, the signal power is $(S_0-(2d-2)S_a)M^2 + {S_a\sum_{j=1}^{d-1} j^2 M^2}/{d^2}$.
% Also, there are $S_a$ new additions and $S_a$ removals at each $t$.
% (set $\Inc_t$) (set $\Dec_t$) (set $\Con_t$)

Signal Model \ref{sigmod2} does not specify a particular generative model. An example of a signal model that satisfies the above assumptions is the following. At each $t$, $S_a$ new elements, randomly selected from ${N_{t-1}}^c$, get added to the support at initial magnitude, $r$, and equally likely sign. Their magnitude keeps increasing gradually, at rate $r$, for a certain amount of time, $d$, after which it becomes constant at $M:=dr$. The sign does not change. Also, at each time, $t$, $S_a$ randomly selected elements out of the ``stable" elements' set (set of elements which have magnitude $M$ at $t-1$), begin to decrease at rate $r$ and this continues until their magnitude becomes zero, i.e. they get removed from the support. This model is specified mathematically in Appendix \ref{generativemodel}. We use this in our simulations. Another possible generative model is: at each time $t$, randomly select $S_a$ out of the $2S_a$ current elements with magnitude $jr$ and increase them, and decrease the other $S_a$ elements. Do this for all $1 \le j \le d-1$.% Do the rest as above.%and then also select the new addition and the new decreasing set as before.

In practice, different elements may have different magnitude increase rates and different stable magnitudes, but to keep notation simple we do not consider that here. Our results can be extended to this case fairly easily.%
% %For studying stability, we need to assume a signal model. W
%A symmetric model is assumed for coefficient removal.

To understand the implications of the assumptions in Signal Model \ref{sigmod2}, we define the following sets.
\bd
Let
\ben
\item $\Dset_{t}(j): = \{i: |x_{t,i}| = jr, \ |x_{t-1,i}| = (j+1)r \}$ denote the set of elements that {\em decrease} from $(j+1)r$ to $jr$ at time, $t$,%
\item $\Iset_{t}(j): = \{i: |x_{t,i}| = jr, \ |x_{t-1,i}| = (j-1)r \}$ denote the set of elements that {\em increase} from $(j-1)r$ to $jr$ at time, $t$,%
\item $\Sset_t(j):= \{i:  0 < |x_{t,i}| < j r \}$ denote the set of {\em small but nonzero} elements, with smallness threshold $jr$.% equal to
\item Clearly,
\ben
\item the newly added set, $\Aset_t:= \Iset_t(1)$, and the newly removed set, $\Rset_t:= \Dset_t(0)$.
\item $|\Iset_{t}(j)|=S_a$, $|\Dset_{t}(j)|=S_a$ and $|\Sset_t(j)| = 2(j-1)S_a$ for all $j$.
\een
\een
\ed

Consider a $1 < j \le d$. From the signal model, it is clear that at any time, $t$, $S_a$ elements enter the small elements' set, $\Sset_t(j)$, from the bottom (set $\Aset_t$) and $S_a$ enter from the top (set $\Dset_{t}(j-1)$). Similarly $S_a$ elements leave  $\Sset_t(j)$ from the bottom  (set $\Rset_t$) and $S_a$ from the top (set $\Iset_{t}(j)$). Thus,%In other words,
\bea
\Sset_t(j) = \Sset_{t-1}(j)  \cup (\Aset_t \cup \Dset_{t}(j-1)) \setminus (\Rset_t \cup \Iset_{t}(j)) \ \ \
\label{sseteq}
\eea
Since the sets $\Aset_t, \Rset_t, \Dset_{t}(j-1),\Iset_{t}(j)$ are mutually disjoint, and since $\Rset_t \subseteq \Sset_{t-1}(j)$ and $\Iset_{t}(j) \subseteq \Sset_{t-1}(j)$, thus,%
\bea
\Sset_{t-1}(j)  \cup \Aset_t  \setminus \Rset_t = \Sset_t(j) \cup \Iset_{t}(j) \setminus \Dset_{t}(j-1)
\label{sseteq_2}
\eea
We will use this in the proof of the stability result of Sec. \ref{addLSdel_modcs}.% given Appendix \ref{proof_addLSdel_modcs}.%

\section{Stability of modified-CS}
\label{simple_modcs}
Modified-CS was first introduced in \cite{modcsjp} as a solution to the problem of sparse reconstruction with partial, and possibly erroneous, knowledge of the support. Denote this ``known" support by $T$. Modified-CS tries to find a signal that is sparsest outside of the set $T$ among all signals satisfying the data constraint. For recursively reconstructing a time sequence of sparse signals, we use the support estimate from the previous time, $\Nhat_{t-1}$ as the set $T$. At the initial time, $t=0$, we let $T$ be the empty set, i.e. we do simple CS\footnote{Alternatively, as explained in \cite{modcsjp}, we can use prior knowledge of the initial signal's support as the set $T$ at $t=0$.}. Thus at $t=0$ we need more measurements, $n_0 > n$. Denote the $n_0 \times m$ measurement matrix used at $t=0$ by $A_0$.% But usually even this is not so accurate and so more measurements are needed at $t=0$

We summarize the modified-CS algorithm in Algorithm \ref{modcsalgo}. Here $\alpha$ denotes the support estimation threshold.

% In this section, we find the conditions under which we can obtain a time-invariant bound on the sizes of these sets, i.e. ensure ``stability". This ensures a time-invariant bound on the reconstruction errors.%will ensure a time-invariant bound on the modified-CS (CS-residual) error and on the final LS error for both methods.%

%In this section, we begin by first summarizing the modified-CS algorithm for recursive reconstruction of signal sequences from noisy measurements. The noiseless measurements version was first proposed in \cite{a}. Next we discuss the key steps leading to the stability result and the result itself.  Finally its implications are discussed.

%\subsection{Modified-CS algorithm}
%As explained in \cite{a}, at the initial time, $t=0$, we either do modified-CS with using a prior knowledge of the support or we do simple CS (equivalent to modified-CS with the ``known support" being an empty set). The prior knowledge is usually not very accurate and thus at $t=0$ one will usually need more measurements, $n_0 > n$. Denote the $n_0 \times m$ measurement matrix used at $t=0$ as $A_0$. As mentioned earlier, $A$ denotes the $n \times m$ measurement matrix used at $t>0$.  For simplicity, in this work, assume that we do simple CS at $t=0$.

\vspace{-0.15in}
\begin{algorithm}[h!]
\caption{{\bf \small Modified-CS}}
%At $t=0$, compute $\hat{x}_{0}$ as the solution of $\min_{\beta}  \|(\beta)\|_1 \ \text{s.t.} \ ||y_0 -A_0 \beta|| \le \eps$. Compute $\Nhat_0 = \{i \in [1,m] : |(\xhat_{0,modcs})_i| > \alpha \}$. \\
For $t \ge 0$, do
\ben
\item {\em Simple CS. } If $t = 0$, set $T = \emptyset$ and compute $\xhat_{t,modcs}$ as the solution of
\bea
\min_\beta  \|(\beta)\|_1 \ \text{s.t.} \ || y_0 - A_0 \beta || \le \eps
\label{simpcs}
\eea

\item {\em Modified-CS. } If $t>0$, set $T = \Nhat_{t-1}$ and compute $\xhat_{t,modcs}$ as the solution of
\label{step1_0}
\bea
\min_\beta  \|(\beta)_{T^c}\|_1 \ \text{s.t.} \ || y_t - A \beta || \le \eps
\label{modcs}
\eea

%\item {\em Modified-CS. } Let $T = \Nhat_{t-1}$. Compute $\xhat_{t,modcs}$ as the solution of
%\bea
%\min_\beta  \|(\beta)_{T^c}\|_1x \ \text{s.t.} \ || y_t - A \beta || \le \eps
%\label{modcs}
%\eea
%\label{step1noiseless}

\item {\em Estimate the Support. } Compute $\tT$ as
\bea
\tT=\{i \in [1,m] : |(\xhat_{t,modcs})_i| > \alpha \}
\eea

\item Set $\Nhat_t = \tT$. Output $\hat{x}_{t,modcs}$. Feedback $\Nhat_t$.%, increment $t$, and go to step \ref{step1noiseless}.
\een
\label{modcsalgo}
\end{algorithm}
\vspace{-0.15in}

By adapting the approach of \cite{candes_rip}, the error of modified-CS can be bounded as a function of $|T|=|N|+|\Delta_e|-|\Delta|$ and $|\Delta|$. This was done in \cite{arxiv}. We state its modified version here.%of \cite{arxiv}'s result.

\begin{lemma}[modified-CS error bound \cite{arxiv}]
If  $\|w\| \le \eps$ and $\delta_{|N|+|\Delta|+|\Delta_e|} < \sqrt{2}-1$, then
\bea
\|x_t - \xhat_{t,modcs}\| \sle C_1(|N|+|\Delta|+|\Delta_e|)) \eps, \ \text{where} \nn \\
C_1(S) \sdefn \frac{4 \sqrt{1+\delta_S}}{1 - (\sqrt{2} +1) \delta_S}
\label{defC1s}
\eea
\label{modcsbnd}
\end{lemma}

If $\delta_{|N|+|\Delta|+|\Delta_e|}$ is just smaller than $\sqrt{2}-1$, the error bound will be very large because the denominator of $C_1(S)$ will be very large. To keep the bound small, we need to assume that it is smaller than $b(\sqrt{2}-1)$ with a $b<1$. For simplicity, let $b=1/2$.  Then we get the following corollary, which we will use in our stability results.
\begin{corollary}[modified-CS error bound]
If  $\|w\| \le \eps$ and $\delta_{|N|+|\Delta|+|\Delta_e|} < (\sqrt{2}-1)/2$, then
\label{modcs_cs_bnd}
\bea
\|x_t - \xhat_{t,modcs}\| \sle C_1(|N|+|\Delta|+|\Delta_e|)) \eps \le 8.79 \eps
\eea
\end{corollary}
Proof: Notice that $C_1(S)$ is an increasing function of $\delta_S$. The above corollary follows by using  $\delta_{|N|+|\Delta|+|\Delta_e|} < (\sqrt{2}-1)/2$ to bound $C_1(S)$.
%
%. %\frac{4 \sqrt{1+0.207}}{1 - (\sqrt{2} +1)(\sqrt{2}-1)/2} \eps \le Notice that $C_1(S)$ is an increasing function of $\delta_S$.
%
Next, we state a similarly modified version of the result for CS \cite{candes_rip}. %that ensures the same error bound as in the previous corollary. This will be referenced for comparison later.
\begin{corollary}[CS error bound \cite{candes_rip}]
If  $\|w\| \le \eps$ and $\delta_{2|N|} < (\sqrt{2}-1)/2$, then
\bea
\|x_t - \xhat_{t,cs}\| \sle C_1(2|N|) \eps \le 8.79 \eps
\eea
where $\xhat_{t,cs}$ is the solution of (\ref{modcs}) with $T = \emptyset$ (empty set).%
\label{cs_bnd}
\end{corollary}

\subsection{Stability result for modified-CS}
\label{stab_modcs}

The first step to show stability is to find sufficient conditions for a certain set of large coefficients to definitely get detected, and for the elements of $\Delta_e$ to definitely get deleted. These can be obtained using Corollary  \ref{modcs_cs_bnd} and the following simple facts which we state as a proposition.%, in order to easily refer to them later.%and (b) to definitely not get falsely deleted, and (c)

\begin{proposition}(simple facts)
\ben
\item An $i \in N$ will definitely get detected if $|x_i|  > \alpha + \|x - \xhat_{modcs}\|$. This follows since  $ \|x - \xhat_{modcs}\| \ge \|x - \xhat_{modcs}\|_\infty \ge |(x - \xhat_{modcs})_{i}|$.%(unknown part of support) definitely
\label{det1}

%\item Similarly, an $i \in T$ will not get falsely deleted if $|x_i|  > \alpha + \|x - \xhat_{modcs}\|$. \label{nofalsedel1}

\item Similarly, all $i \in \Delta_{e}$ (the zero elements of $T$) will definitely get deleted if $\alpha \ge \|x - \xhat_{modcs}\|$.

%\item The bound in fact \ref{errls1} is non-decreasing in $|\tT_\dett|$ and $|\tDelta_\dett|$.\label{nondec}
\een
\label{prop0}
\end{proposition}

Combining the above facts with Corollary \ref{modcs_cs_bnd}, we get the following lemma.
\begin{lemma} %[Detection condition]
Assume that $\|w\| \le \eps$, $|N| \le S_N$, $|\Delta_e| \le S_{\Delta_e}$, $|\Delta| \le S_\Delta$. %Also assume that
\ben
\item Let $L:=\{i \in N: |x_i| \ge b_1 \}$. All elements of $L$ will get detected at the current time if $\delta_{S_N + S_{\Delta_e} + S_\Delta} < (\sqrt{2}-1)/2$ and $b_1 > \alpha +  8.79 \eps$.

\item There will be no false additions, and all the true removals from the support (the set $\Delta_{e,t}$) will get deleted at the current time, if $\delta_{S_N + S_{\Delta_e} + S_\Delta} < (\sqrt{2}-1)/2$ and $\alpha \ge 8.79 \eps$. %C_1(S_N + S_{\Delta_e} + S_\Delta) \eps %8.79 \eps %

\een
\label{lemma_modcs}
\end{lemma}
Notice that in the above lemma and proposition, for ease of notation, we have removed the subscript $t$.

We use the above lemma to obtain the stability result as follows. Let us fix a bound on the maximum allowed magnitude of a missed coefficient.
Suppose we want to ensure that only coefficients with magnitude less than $2r$ are part of the final set of misses, $\tDelta_t$, at any time, $t$ and that the final set of extras, $\tDelta_{e,t}$ is an empty set. In other words, we find conditions to ensure that $\tDelta_t \subseteq \Sset_{t}(2)$ (using Signal Model \ref{sigmod2}, this will imply that $|\tDelta_t| \le 2S_a$) and $|\tDelta_{e,t}|=0$. This leads to the following result. The result can be easily generalized to ensure that $\Delta_t \subseteq \Sset_t(d_0)$, and thus $|\Delta_t| \le (2d_0-2)S_a$, holds at all times $t$, for some $d_0 \le d$ (what we state below is the $d_0=2$ case). %That will weaken the set of assumptions required.

\begin{theorem}[Stability of modified-CS]
Assume Signal Model \ref{sigmod2} and $\|w_t\| \le \eps$. If the following hold
\ben

\item {\em (support estimation threshold) } set $\alpha =   8.79 \eps$ %C_1(S_0 + 3S_a) \eps$
\label{threshes_simple}

\item {\em (support size, support change rate)} $S_0,S_a$ satisfy $\delta_{S_0 + 3S_a} < (\sqrt{2}-1)/2$ ,
\label{measmodel_simple}

\item {\em (new element increase rate) } $r \ge G$, where
\label{add_del_simple}
\bea
G \sdefn \frac{ \alpha + 8.79 \eps }{2} = 8.79 \eps
%G \sdefn \frac{ \alpha + C_1(S_0 + 3S_a) \eps }{2} = C_1(S_0 + 3S_a) \eps \le 8.79 \eps
\eea

\item {\em (initial time)} at $t=0$, $n_0$ is large enough to ensure that $\tDelta_t  \subseteq \Sset_0(2)$, $|\tDelta_t| \le 2S_a$,  $|\tDelta_{e,t}| =0$ and $|\tT_t| \le S_0$
\label{initass_simple}
\een
then, at all $t \ge 0$,
\ben
\item  $|\tT_t| \le S_0$, $|\tDelta_{e,t}| =0$, $\tDelta_t \subseteq \Sset_t(2)$ and so $|\tDelta_t| \le 2S_a$,
\item $|T_t| \le S_0$, $|\Delta_{e,t}| \le S_a$, and $|\Delta_t| \le 2S_a$,
\item $\|x_t - \xhat_{t,modcs}\| \le 8.79 \eps$ %\le C_1(S_0+3S_a) \eps
\een
\label{stabres_simple_modcs}
\end{theorem}

{\em Proof: } The proof follows using induction. We use the induction assumption; $T_t = \tT_{t-1}$; and the fact that $N_t = N_{t-1} \cup \Aset_t \setminus \Rset_t$ to bound $|T_t|$, $|\Delta_t|$ and $|\Delta_{e,t}|$. Next, we use these bounds and Lemma \ref{lemma_modcs} to bound $|\tDelta_t|$ and $|\tDelta_{e,t}|$. Finally we use $|\tT_t| \le |N_t| +  |\tDelta_{e,t}|$ to bound $|\tT_t|$. The complete proof is given in Appendix \ref{proof_simple_modcs}.

\subsection{Discussion}
First notice that condition \ref{initass} is not restrictive. It is easy to see that this will hold if the number of measurements at $t=0$, $n_0$, is large enough to ensure that $A_0$ satisfies $\delta_{2S_0} \le (\sqrt{2}-1)/2$.% and the addition and deletion thresholds satisfy condition \ref{threshes_simple}. %and the magnitude increase/decrease rate, $r$ satisfies condition \ref{add_del_simple}

%While the above result looks complicated, all it says is the following. For a given number and type of measurements (i.e. for a given measurement matrix, $A$), and for a given maximum noise level, $\eps$, if,
%\ben
%\item the support estimation threshold is large enough to satisfy condition \ref{threshes_simple},
%\item the support size and the support change size are small enough to satisfy condition \ref{measmodel_simple},
%\item the newly added coefficients increase (existing large coefficients decrease) at least at rate $r$ which satisfies condition \ref{add_del_simple}, and
%\item the number of measurements at $t=0$, $n_0$, is large enough so that condition \ref{initass_simple} holds,
%\een
%then the support errors are bounded by time-invariant values, $2S_a$ and zero. Consequently the reconstruction error is also bounded by a time-invariant value.

Clearly, when $S_a \ll S_0$ (slow support change), the support error bound of $2S_a$ is small compared to support size, $S_0$, making it {\em a meaningful} stability result.

Compare the maximum allowed support size $S_0$ that is needed for stability of modified-CS with what simple CS needs. Since simple CS is not a recursive approach (each time instant is handled separately), Corollary \ref{cs_bnd}, also serves as a stability result for simple CS.
From Corollary \ref{cs_bnd}, CS needs $\delta_{2S_0} < (\sqrt{2}-1)/2$ to ensure that its error is bounded by $8.79 \eps$ for all $t$. On the other hand, for $t>0$, our result from Theorem \ref{stabres_simple_modcs} only needs $\delta_{S_0 + 3S_a} < (\sqrt{2}-1)/2$ to get the same error bound, while, at $t=0$, it needs the same condition as CS. Said another way, for a given $S_0$,  at $t=0$, we need as many measurements as CS does\footnote{This can also be improved if we use prior support knowledge at $t=0$ as explained in \cite{modcsjp}.}, while at $t>0$, we can use much fewer measurements, only enough to satisfy $\delta_{S_0 + 3S_a} < (\sqrt{2}-1)/2$.  When $S_a \ll S_0$ (slow support change), this is clearly much weaker.

\subsection{Limitations}
\label{limitations}
We now discuss the limitations of the above result and of modified-CS. First, in Proposition \ref{prop0} and hence everywhere after that we bound the $\ell_\infty$ norm of the error by the $\ell_2$ norm. This is clearly a loose bound and results in a loose lower bound on the required threshold $\alpha$ and consequently a larger than required lower bound on the minimum required rate of coefficient increase/decrease, $r$.

Second, we use a single threshold $\alpha$ for addition and deletion to the support estimate. To ensure deletion of the extras, we need $\alpha$ to be large enough. But this means that $r$ needs to be even larger to ensure correct detection (and no false deletion) of all but the smallest $2S_a$ elements.
There is another related issue which is not seen in the theoretical analysis because we only bound $\ell_2$ norm of the error, but is actually more important since it affects reconstruction itself, not just the sufficient conditions for its stability. This has to do with the fact that $\xhat_{t,modcs}$ is a biased estimate of $x_t$. A similar issue for noisy CS, and a possible solution (Gauss-Dantzig selector), was first discussed in \cite{dantzig}. In our context, along $T^c$, the values of $\xhat_{t,modcs}$ will be biased towards zero, while along $T$ they may be biased away from zero (since there is no constraint on $(\beta)_T$). The bias will be larger when the noise is larger. This will create the following problem. The set $T$ contains the set $\Delta_e$ which needs to be deleted. Since the estimates along $\Delta_e$ may be biased away from zero, one will need a higher threshold to delete them. But that would make detecting new additions more difficult, especially since the estimates along $\Delta \subseteq T^c$ are biased towards zero.%

\section{Stability of Modified-CS with Add-LS-Del}
\label{addLSdel_modcs}

The last two issues mentioned above in Sec. \ref{limitations} can be partly addressed by replacing the single support estimation step by a support addition step (that uses a smaller threshold), followed by an LS estimation step and then a deletion step that thresholds the LS estimate. The addition step threshold needs to be just large enough to ensure that the matrix used for LS estimation is well-conditioned. If the threshold is chosen properly and if $n$ is large enough, the LS estimate will have smaller error than the modified-CS output. As a result, deletion will be more accurate and in many cases one can also use a larger deletion threshold. The addition-LS-deletion idea was simultaneously introduced in \cite{cosamp} (CoSaMP) for a static sparse reconstruction and in our older work \cite{just_lscs,kfcsicip} (LS-CS and KF-CS) for recursive reconstruction of sparse signal sequences.% using a greedy approach

%\subsection{Modified-CS with Add-LS-Del Algorithm}
Let $\alpha_{add}$ denote the addition threshold and let $\alpha_{del}$ denote the deletion threshold. We summarize the algorithm in Algorithm \ref{modcsalgo_2}.

\begin{algorithm}[h]
\caption{{\bf \small Modified-CS with Add-LS-Del}}
%At $t=0$, compute $\hat{x}_{0}$ as the solution of $\min_{\beta}  \|(\beta)\|_1 \ \text{s.t.} \ ||y_0 -A_0 \beta|| \le \eps$. Compute $\Nhat_0 = \{i \in [1,m] : |(\xhat_{0,modcs})_i| > \alpha \}$. \\
For $t \ge 0$, do
\ben
\item {\em Simple CS. } If $t = 0$, set $T = \emptyset$ and compute $\xhat_{t,modcs}$ as the solution of (\ref{simpcs}).
%\bea
%\min_\beta  \|(\beta)\|_1 \ \text{s.t.} \ || y_0 - A_0 \beta || \le \eps
%\label{simpcs}
%\eea

\item {\em Modified-CS. } If $t>0$, set $T = \Nhat_{t-1}$ and compute $\xhat_{t,modcs}$ as the solution of of (\ref{modcs}).
%\bea
%\min_\beta  \|(\beta)_{T^c}\|_1 \ \text{s.t.} \ || y_t - A \beta || \le \eps
%\label{modcs}
%\eea
\label{step1}

\item {\em Additions / LS.} Compute $T_\dett$ and LS estimate using it:%
\label{addls}
\bea
T_\dett \se T \cup \{i \in T^c: |(\xhat_{t,modcs})_i| > \alpha_{add} \} \nn \\
(\xhat_{t,\dett})_{T_\dett} \se {A_{T_\dett}}^\dag y_t, \ \ (\xhat_{t,\dett})_{T_\dett^c} = 0
\label{Tdett}
\eea

\item {\em Deletions / LS.} Compute $\tT$ and LS estimate using it:%
\label{delete}
\bea
\tT \se  T_{\dett} \setminus  \{i \in T_\dett: |(\xhat_{t,\dett})_i| \le \alpha_{del} \} \nn \\
%\eea
%Compute the LS estimate using $\tT$:
%\bea
(\xhat_{t})_{\tT} \se {A_{\tT}}^\dag  y_t, \ \ (\xhat_{t})_{\tT^c} = 0  %{A_{\Nhat_t}}^\ddag
\label{finalls}
\eea

\item Set $\Nhat_t = \tT$. Output $\xhat_t$. Feedback $\Nhat_t$. %Increment $t$. %Go to step \ref{step1}.

\een
%Feedback $\Nhat_t$, increment $t$, and go to step \ref{step1noiseless}.
%\vspace{-0.05in}
\label{modcsalgo_2}
\end{algorithm}

\bd [Define $T_\dett,\Delta_\dett, \Delta_{e,\dett}$]
The set $T_\dett$ is the set obtained after the support addition step. It is defined in (\ref{Tdett}) in Algorithm \ref{modcsalgo_2}. We use $\Delta_\dett:=N_t \setminus  T_\dett$ to denote the missing elements from $T_\dett$ and we use $\Delta_{e,\dett}:=T_\dett \setminus N_t$ to denote the extras.% in it.
\ed

\subsection{Stability result for modified-CS with add-LS-del}
\label{stab_modcs}
%Our approach is as follows. We use Lemma \ref{modcsbnd} to obtain conditions for an element of the unknown support, $\Delta$, to definitely get detected at $t$. Next, we use a bound on the error in the LS step after the detection step to obtain conditions for a nonzero element of $\tT_{\dett}$ to not get falsely deleted and for all elements of $\Delta_{e,\dett}$ to get deleted. The following simple facts do not require a proof.

The first step to show stability is to find sufficient conditions for (a) a certain set of large coefficients to definitely get detected, and (b) to definitely not get falsely deleted, and (c) for the zero coefficients in $T_\dett$ to definitely get deleted. These can be obtained using  Corollary \ref{modcs_cs_bnd} and the following simple facts which we state as a proposition, in order to easily refer to them later.
\begin{proposition}(simple facts)
\ben
\item An $i \in \Delta$ will definitely get detected if $|x_i|  > \alpha_{add} + \|x - \xhat_{modcs}\|$. This follows since  $ \|x - \xhat_{modcs}\| \ge \|x - \xhat_{modcs}\|_\infty \ge |(x - \xhat_{modcs})_{i}|$.%(unknown part of support) definitely
\label{det1}

\item Similarly, an $i \in T_\dett$ will definitely not get falsely deleted if $|x_i| >  \alpha_{del} + \|(x - \xhat_\dett)_{T_\dett}\|$.% definitely(the nonzero elements of $T_\dett$) (T_\dett \setminus \Delta_{e,\dett})
\label{nofalsedel1}

\item All $i \in \Delta_{e,\dett}$ (the zero elements of $T_\dett$) will definitely get deleted if $\alpha_{del} \ge \|(x - \xhat_\dett)_{T_\dett}\|$.% \ge |(\xhat_\dett)_i|$.% definitely
\label{truedel1}

\item Consider LS estimation using known part of support $T$, i.e. consider the estimate $(\xhat_{LS})_T = {A_T}^\dag y$ and $(\xhat_{LS})_{T^c} = 0$ computed from $y:=Ax+w$. Let $\Delta = N \setminus T$ where $N$ is the support of $x$. If $\|w\| \le \eps$ and if $\delta_{|T|} < 1/2$, then $\|(x - \xhat_{LS})_{T}\| \le \sqrt{2} \eps + 2{\theta_{|T|,|\Delta|}} \|x_{\Delta}\|$ (instead of $\delta_{|T|} < 1/2$, one can pick any $b<1$ and the constants in the bound will change appropriately). This bound is derived in \cite[equation (15)]{just_lscs}.
\label{errls1}
%If $\|w\| \le \eps$ and if $\delta_{|T_\dett|} < 1/2$ (or pick any $b<1$ and the constants will change appropriately), then $\|(x - \xhat_\dett)_{T_\dett}\| \le \sqrt{2} \eps + 2{\theta_{|T_\dett|,|\Delta_\dett|}} \|x_{\Delta_\dett}\|$.\label{errls1}

%\item The bound in fact \ref{errls1} is non-decreasing in $|T_\dett|$ and $|\Delta_\dett|$.\label{nondec}
\een
\label{prop1}
\end{proposition}

Combining the above facts with Corollary \ref{modcs_cs_bnd}, we can easily get the following three lemmas.

\begin{lemma}[Detection condition]
Assume that $\|w\| \le \eps$, $|N| \le S_N$, $|\Delta_e| \le S_{\Delta_e}$, $|\Delta| \le S_\Delta$. Let $\Delta_1:=\{i \in \Delta: |x_i| \ge b_1 \}$. All elements of $\Delta_1$ will get detected at the current time if $\delta_{S_N + S_{\Delta_e} + S_\Delta} < (\sqrt{2}-1)/2$ and  $b_1 > \alpha_{add} + 8.79\eps$.
 %C_1(S_N + S_{\Delta_e} + S_\Delta) \eps  %S_N + S_{\Delta_e} + S_\Delta) \eps
\label{detectcond_modcs}
\end{lemma}

{\em Proof: } The lemma follows from fact \ref{det1} of Proposition \ref{prop1} and Corollary \ref{modcs_cs_bnd}.% and the fact that $C_1(.)$ is a non-decreasing function of $|N|,|\Delta|,|\Delta_e|$.

\begin{lemma}[No false deletion condition]
Assume that $\|w\| \le \eps$, $|T_\dett| \le S_T$ and $|\Delta_\dett| \le S_\Delta$. For a given $b_1$, let $T_1:=\{i \in T_\dett: |x_i| \ge b_1\}$. All $i \in T_1$ will not get (falsely) deleted at the current time if $\delta_{S_T} < 1/2$ and $b_1 > \alpha_{del} + \sqrt{2} \eps + 2{\theta_{S_T,S_\Delta}} \|x_{\Delta_\dett}\|$
\label{nofalsedelscond}
\end{lemma}

{\em Proof: } This follows directly from fact \ref{nofalsedel1} and fact \ref{errls1} (applied with $T \equiv T_\dett$ and $\Delta \equiv \Delta_\dett$) of Proposition \ref{prop1}.% and \ref{nondec}.

\begin{lemma}[Deletion condition]
Assume that $\|w\| \le \eps$, $|T_\dett| \le S_T$ and $|\Delta_\dett| \le S_\Delta$. All elements of $\Delta_{e,\dett}$ will get deleted if $\delta_{S_T} < 1/2$ and $\alpha_{del} \ge \sqrt{2} \eps + 2{\theta_{S_T,S_\Delta}} \|x_{\Delta_\dett}\|$.
\label{truedelscond}
\end{lemma}

{\em Proof: } This follows directly from fact \ref{truedel1} and fact \ref{errls1} (applied with $T \equiv T_\dett$ and $\Delta \equiv \Delta_\dett$) of Proposition \ref{prop1}.% and \ref{nondec}.

Using the above lemmas and the signal model, we can obtain sufficient conditions to ensure that, for some $d_0 \le d$, at each time $t$, $\tDelta_t \subseteq \Sset_t(d_0)$ (so that $|\tDelta| \le (2d_0-2)S_a$) and $|\tDelta_{e,t}|=0$, i.e. only elements smaller than $d_0 r$ may be missed and there are no extras. For notational simplicity, we state the special case below which uses $d_0=2$. The general case is given in Appendix \ref{stabres_modcs_gen}.

%We state here a special case of Theorem \ref{stabres_modcs} which follows by setting $d_0=2$ and maximum allowed false detects, $f=S_a$.

\begin{theorem}[Stability of modified-CS with add-LS-del]
Assume Signal Model \ref{sigmod2} and  $\|w_t\| \le \eps$. If% the following hold
\ben
\item {\em (addition and deletion thresholds) }
\ben
\item $\alpha_{add}$ is large enough so that there are at most $S_a$ false additions per unit time,
\label{addthresh}

\item $\alpha_{del}  = \sqrt{2} \eps +  2 \sqrt{S_a} \theta_{S_0+2S_a,S_a} r $,
\label{delthresh}
\een
\label{add_del_thresh}

\item {\em (support size, support change rate)} $S_0,S_a$ satisfy
\ben
%\item $\delta_{S_0+2S_a} < 1/2$,%$k^a S_a \le \sinf$, $S_0+S_a + f \le \sone$,
\item $\delta_{S_0 + 3S_a} < (\sqrt{2}-1)/2$,
%\item $\theta_{S_0+2S_a,S_a} < \frac{1}{2} \sqrt{  \frac{1}{8S_a } }$, %\sum_{j=1}^{d_0-1} j^2 }
\item $\theta_{S_0+2S_a,S_a} < \frac{1}{2}  \frac{1}{2\sqrt{S_a}}$, %\sum_{j=1}^{d_0-1} j^2 }
\label{theta_ass}
\een
\label{measmodel}

\item {\em (new element increase rate) } $r \ge \max(G_1,G_2)$, where
\label{add_del}
\bea
G_1 \sdefn  \frac{ \alpha_{add} + 8.79 \eps }{2} \nn \\ %\frac{ \alpha_{add} + C_1(S_0 + 3S_a ) \eps }{2} \le
G_2 \sdefn \frac{\sqrt{2} \eps}{1 - 2\sqrt{S_a }\theta_{S_0+2S_a,S_a}}  %\le 2\sqrt{2} \eps   \ \ \ \ \ \ \
\eea
\item {\em (initial time)} at $t=0$, $n_0$ is large enough to ensure that $\tDelta_t  \subseteq \Sset_0(2)$, $|\tDelta| \le 2S_a$,  $|\tDelta_{e,t}| =0$, $|\tT_t| \le S_0$,%$|\tDelta| \le (2d_0-2)S_a$, we use enough measurements,
\label{initass}
\een
then,  at all $t \ge 0$,
\ben
\item $|\tT_t| \le S_0$, $|\tDelta_{e,t}| =0$, and $|\tDelta_t| \le 2S_a$, %and $\tDelta \subseteq \Sset_t(2)$ and so

\item $|T_t| \le S_0$, $|\Delta_{e,t}| \le S_a$, and $|\Delta_t| \le 2S_a$

\item $|\tT_{\dett,t}| \le S_0+2S_a$, $|\tDelta_{e,\dett,t}| \le 2S_a$, and $|\tDelta_{\dett,t}| \le S_a$ %at all $t > 0$,

\item $\|x_t-\xhat_t\| \le \sqrt{2} \eps + (2\theta_{S_0,2S_a}+1) \sqrt{2S_a} r$

\item $\|x_t - \xhat_{t,modcs}\| \le 8.79 \eps$ %\le C_1(S_0+3S_a) \eps
\een
\label{stabres_modcs}
\end{theorem}

{\em Proof: }
The proof again follows by induction, but is more complicated than that in the previous section, due to the support addition and deletion steps.
The induction step consists of three parts. First, we use the induction assumption; $T_t = \tT_{t-1}$; and the fact that $N_t = N_{t-1} \cup \Aset_t \setminus \Rset_t$ to bound $|T_t|,|\Delta_{e,t}|,|\Delta_t|$. This part of the proof is the same as that of Theorem \ref{stabres_simple_modcs}. The next two parts are different and more complicated. We use the bounds from the first part; equation (\ref{sseteq_2}); Lemma \ref{detectcond_modcs}; the limit on the number of false detections; and $|T_\dett| \le |N| + |\Delta_{e,\dett}|$ to bound $|\Delta_{\dett,t}|,|\Delta_{e,\dett,t}|,|T_{\dett,t}|$. Finally, we use the bounds from the second part and Lemmas \ref{nofalsedelscond} and \ref{truedelscond} to bound $|\tDelta_t|,|\tDelta_{e,t}|,|\tT_t|$.  The complete proof is given in Appendix \ref{proof_addLSdel_modcs}.

\subsection{Discussion} % (so that its $\infty$ norm is significantly smaller than its $\ell_2$ norm),
Notice that condition \ref{theta_ass} may become difficult to satisfy as soon as $S_a$ increases, which will happen when the problem dimension, $m$, increases (and consequently $S_0$ increases, even though $S_a$ and $S_0$ remain small fractions of $m$, e.g. typically $S_0 \approx 10\%m$ and $S_a \approx 0.2\%m$). The reason we get this condition is because in facts \ref{nofalsedel1} and \ref{truedel1} of Proposition \ref{prop1}, and hence also in Lemmas \ref{nofalsedelscond} and \ref{truedelscond} and the final result, we bound the $\ell_\infty$ norm of the LS step error, $\|(x - \xhat_\dett)_{T_\dett}\|_\infty$ by the $\ell_2$ norm, $\|(x - \xhat_\dett)_{T_\dett}\|$.

This is clearly a loose bound - it holds with equality only when the entire LS step error is concentrated in one dimension. In practice, as observed in simulations, this is usually not true. The LS step error is quite spread out since the LS step tends to reduce the bias in the estimate. Thus it is not unreasonable to assume that $\|(x - \xhat_\dett)_{T_\dett}\|_\infty \le \|(x - \xhat_\dett)_{T_\dett}\| / \sqrt{S_a}$ (LS step error is spread out enough to ensure this) at all times. In simulations, we observed that when $m=200$, $S_0=20$, $S_a=2$,  $r=3/4$, $n=59$, $w_t \sim^{i.i.d.} unif(-c,c)$ with $c=0.1266$, and we used $\alpha_{add}=c/2$, $\alpha_{del}=r/2$ this was true 99.8\% of the times. The same was true even when $r$ was reduced to $2/5$. When we increased the problem size five times to $m=1000$, $S_0=100$, $S_a=10$, $n=295$, and all other parameters were the same, this was true 93\% of the times. In all cases, 100\% of the times, $(\|(x - \xhat_\dett)_{T_\dett}\|/\sqrt{S_a}) /\|(x - \xhat_\dett)_{T_\dett}\|_\infty < 0.78$. All simulations used a random-Gaussian matrix $A$.

With this extra assumption, Lemmas \ref{nofalsedelscond} and \ref{truedelscond} will get replaced by the following two lemmas. With using these new lemmas, condition \ref{theta_ass} will get replaced by $\theta_{S_0+2S_a,S_a} < 1/4$ which is an easily satisfiable condition. Moreover, this also makes the lower bound on the required value of $r$ (rate of coefficient increase/decrease) smaller.

\begin{lemma}
Assume that $\|w\| \le \eps$, $|T_\dett| \le S_T$ and $|\Delta_\dett| \le S_\Delta$. Also, assume that $\|(x - \xhat_\dett)_{T_\dett}\|_\infty \le \|(x - \xhat_\dett)_{T_\dett}\| / \sqrt{S_a}$ (the LS step error is spread out enough). For a given $b_1$, let $T_1:=\{i \in T_\dett: |x_i| \ge b_1\}$. All $i \in T_1$ will not get (falsely) deleted at the current time if  $\delta_{S_T} < 1/2$, and $b_1 > \alpha_{del} + (\sqrt{2} \eps + 2{\theta_{S_T,S_\Delta}} \|x_{\Delta_\dett}\|)/ \sqrt{S_a}$.%
%
%\item  $\|(x - \xhat_\dett)_{T_\dett}\|_\infty = \|(x - \xhat_\dett)_{T_\dett}\| / \sqrt{|T_\dett|}$ (the LS step error is equal in all directions), and
%\item $b_1 > \alpha_{del} + (\sqrt{2} \eps + 2{\theta_{S_T,S_\Delta}} \|x_{\Delta_\dett}\|)/ \sqrt{|T_\dett|}$. %Since $|T_\dett| \ge |N| - |\Delta| \ge  S_0 - S_\Delta$
\label{nofalsedels_equalLS}
\end{lemma}

\begin{lemma}[Deletion condition]
Assume that $\|w\| \le \eps$, $|T_\dett| \le S_T$ and $|\Delta_\dett| \le S_\Delta$. Also, assume that $\|(x - \xhat_\dett)_{T_\dett}\|_\infty \le \|(x - \xhat_\dett)_{T_\dett}\| / \sqrt{S_a}$ (the LS step error is spread out enough). All elements of $\Delta_{e,\dett}$ will get deleted if $\delta_{S_T} < 1/2$ and $\alpha_{del} \ge (\sqrt{2} \eps + 2{\theta_{S_T,S_\Delta}} \|x_{\Delta_\dett}\|)/ \sqrt{S_a}$.
%\item  $\|(x - \xhat_\dett)_{T_\dett}\|_\infty = \|(x - \xhat_\dett)_{T_\dett}\| / \sqrt{|T_\dett|}$ (the LS step error is equal in all directions), and
%\item $\alpha_{del} \ge (\sqrt{2} \eps + 2{\theta_{S_T,S_\Delta}} \|x_{\Delta_\dett}\|)/ \sqrt{|T_\dett|}$.
%
\label{truedelscond_equalLS}
\end{lemma}

%With using these two lemmas, condition \ref{theta_ass} will get replaced by $\theta_{S_0+2S_a,S_a} < 1/4$ which is an easily  satisfiable condition. Moreover, this will also make the smallest required $\alpha_{del}$ smaller, as well as make $G_2$ smaller. Smaller $G_2$ will ensure that a smaller rate of coefficient increase/decrease, $r$, suffices.

By using Lemmas \ref{nofalsedels_equalLS} and \ref{truedelscond_equalLS} instead of Lemmas \ref{nofalsedelscond} and \ref{truedelscond} respectively,  and doing everything else exactly as in the proof of Theorem \ref{stabres_modcs}, we get the following corollary. %Notice that condition \ref{theta_ass} will get replaced by $\theta_{S_0+2S_a,S_a} < 1/4$ which is an easily satisfiable condition even for large sized problems. Moreover, this also makes the smallest required $\alpha_{del}$ smaller, as well as makes $G_2$ smaller. Smaller $G_2$ ensures that a smaller rate of coefficient increase/decrease, $r$, suffices.

%and by using $|T_\dett| \ge |N| - |\Delta_\dett| \ge S_0 - S_\Delta$ to upper bound $1/|T_\dett|$,

\begin{corollary}[Stability of modified-CS with add-LS-del]
Assume Signal Model \ref{sigmod2} and  $\|w_t\| \le \eps$. Let $e_t:=(x_t - \xhat_{\dett,t})_{T_{\dett,t}}$. Assume that $\|e_t\|_\infty \le \|e_t\| / \sqrt{S_a}$ at all $t$ (the LS step error is spread out enough). If
%Also assume that $\|(x - \xhat_\dett)_{T_\dett}\|_\infty \le \|(x - \xhat_\dett)_{T_\dett}\| / \sqrt{S_a}$ (the LS step error is spread out enough).
%Also assume that $\|(x - \xhat_\dett)_{T_\dett}\|_\infty = \|(x - \xhat_\dett)_{T_\dett}\| / \sqrt{|T_\dett|}$ (the LS step error is equal in all directions). If
\ben
\item {\em (addition and deletion thresholds) }
\ben
\item $\alpha_{add}$ is large enough so that there are at most $S_a$ false additions per unit time,
\label{addthresh}

%\item $\alpha_{del}  = \sqrt{\frac{2}{S_0-S_a}} \eps +  \sqrt{\frac{8S_a}{S_0-S_a}} \theta_{S_0+2S_a,S_a} r $,
\item $\alpha_{del}  = \sqrt{\frac{2}{S_a}} \eps + 2 \theta_{S_0+2S_a,S_a} r $, %\frac{8S_a}{S_a}
\label{delthresh}
\een
\label{add_del_thresh}

\item {\em (support size, support change rate)} $S_0,S_a$ satisfy
\ben
\item  $\delta_{S_0 + 3S_a} \le (\sqrt{2}-1)/2$,
\label{measmod_delta}

%\item $\theta_{S_0+2S_a,S_a} \le  \frac{1}{2} \sqrt{ \frac{S_0-S_a}{8S_a} }$
\item $\theta_{S_0+2S_a,S_a} \le  \frac{1}{4}$
\label{measmod_theta}
\een
\label{measmodel} % $S_a < S_0/4$,

\item {\em (new element increase rate) } $r \ge \max(G_1,G_2)$, where
\label{add_del}
\bea
G_1 \sdefn  \frac{ \alpha_{add} + 8.79 \eps }{2}  \nn \\ %\frac{ \alpha_{add} + C_1(S_0 + 3S_a ) \eps }{2} \le
G_2 \sdefn \frac{\sqrt{2} \eps}{\sqrt{S_a} (1 - 2\theta_{S_0+2S_a,S_a})} %\frac{8S_a}{S_a}
%G_2 \sdefn \frac{\sqrt{2} \eps}{\sqrt{S_0-S_a} (1 - \theta_{S_0+2S_a,S_a}\sqrt{ \frac{8S_a}{S_0-S_a}})}
\eea
%\le  2\sqrt{2} \eps \ \ \ \eea

\item {\em (initial time)} at $t=0$, $n_0$ is large enough to ensure that $\tDelta  \subseteq \Sset_0(2)$, $|\tDelta| \le 2S_a$,  $|\tDelta_e| =0$, $|\tT| \le S_0$,%$|\tDelta| \le (2d_0-2)S_a$, we use enough measurements,
\label{initass}

%\item  $\|(x - \xhat_\dett)_{T_\dett}\|_\infty = \|(x - \xhat_\dett)_{T_\dett}\| / \sqrt{|T_\dett|}$ (the LS step error is equal in all directions)
\een
then all conclusions of Theorem  \ref{stabres_modcs} hold.
\label{cor2_relax}
\end{corollary}
%{\em Proof: } The proof is almost exactly the same as that of Theorem \ref{stabres_modcs} given in Appendix \ref{proof_addLSdel_modcs}, with the difference that  lemmas \ref{nofalsedels_equalLS} and \ref{truedelscond_equalLS} replace lemmas \ref{nofalsedelscond} and \ref{truedelscond} respectively, and we lower bound $|T_\dett| \ge |N| - |\Delta_\dett| \ge S_0 - S_\Delta$.

%Everything else is the same.
Notice that conditions \ref{delthresh}, \ref{measmod_theta} and \ref{add_del} ($r \ge G_2$) are weaker compared to those in Theorem \ref{stabres_modcs}, while others are the same. But of course we also need the LS error is spread out enough.
For large sized problems, condition \ref{measmod_theta} of Theorem \ref{stabres_modcs} is the stronger condition out of the two conditions that $S_0, S_a$ need to satisfy. On the other hand, in this corollary, condition \ref{measmod_delta} is the stronger of the two since its RHS is larger  ($\delta_{S_0 + 3S_a} > \theta_{S_0+2S_a,S_a}$) and its LHS is smaller ($(\sqrt{2}-1)/2=0.207 < 1/4$). Condition \ref{measmod_delta} is easy to satisfy even for large sized problems.

Let us compare our result with the CS result given in Corollary \ref{cs_bnd}. It needs $\delta_{2S_0} < (\sqrt{2}-1)/2=0.207$ to achieve the same error bound as our result. On the other hand, if the LS step error is spread out enough, we only need $\delta_{S_0 + 3S_a} \le (\sqrt{2}-1)/2=0.207$ (this is the stronger of the two conditions on $S_0,S_a$). When $S_a \ll S_0$ (slow support change), in fact as long as $S_a < S_0/3$, this is weaker than what CS needs.% to achieve the same error bound.
%and (b) $\theta_{S_0+2S_a,S_a} < \frac{1}{4}$. Notice that the LHS of (b) is not larger than that of (a) while its RHS is larger. Thus, (a) is the stronger condition.

%This result is again a ``meaningful" result since it again bounds the support errors by $2S_a$ and zero, both of which are small compared to $S_0$ under the slow support change assumption.

% Both are clearly weaker when $S_a \ll S_0$. For example, if $S_a = 0.1 S_0$, then (a) becomes $\delta_{1.3S_0} < (\sqrt{2}-1)/2 = 0.207$, while (b) becomes $\theta_{1.2S_0,0.2S_0} < 0.25$. Now, LHS of (b) is not larger than that of (a) and its RHS is larger. Clearly, (a) is weaker than what CS needs to get the same error bound: $\delta_{2S_0} < (\sqrt{2}-1)/2 = 0.207$.

Finally, let us compare this result with that for modified-CS (without add-LS-del) given in Theorem \ref{stabres_simple_modcs}. Because of add-LS-del, the addition threshold, $\alpha_{add}$, can now be much smaller, as long as the number of false adds is small\footnote{e.g. in simulations with $m=200$, $S_0=20$, $S_a=2$,  $r=0.4$, $n=59$, $w_t \sim^{i.i.d.} unif(-c,c)$ with $c=0.1266$, $\alpha_{add}=0.06$, $\alpha_{del}=r/2$, we were able to use $\alpha_{add}=c/2=0.06$ and still ensure number of false adds less than $S_a$.}. If $\alpha_{add}$ is close to zero, the value of $G_1$ is almost half that of $G$, i.e. the minimum required coefficient increase rate, $r$, gets reduced by almost half. Notice that since $\theta_{S_0+2S_a} \le 1/4$, so $G_2 \le (2\sqrt{2} \eps)/\sqrt{S_a} \le 4.4 \eps \le G_1$, i.e. the upper bound on $G_2$ is smaller than $4.4 \eps$, which is the lower bound on $G_1$. Thus $G_1$ is what decides the minimum allowed value of $r$.%\footnote{Moreover the upper bound of $G_2$ will also not be reached when $S_a \ll S_0$ since, as argued above, when $S_a \ll S_0$, condition \ref{measmod_delta} is the stronger condition.} with direct thresholding for support estimation

\begin{remark}
In the discussion in this paper we have used the special case stability results where we find conditions to ensure that the misses remain below $2S_a$. If we look at the general form of the result, e.g. see Appendix \ref{stabres_modcs_gen} for modified-CS with add-LS-del, the rate of coefficient increase decides what support error level the algorithm stabilizes to, and this, in turn, decides  what conditions on $\delta$ and $\theta$ are needed (in other words, how many measurements, $n$, are needed). For a given $n$, as $r$ is reduced, the algorithm stabilizes to larger and larger support error levels and finally becomes unstable. See Fig. \ref{fig1}. Also, if $n$ is increased, stability can be ensured for smaller $r$'s.%, the others can be generalized similarly)
\end{remark}

\section{Stability of LS-CS (CS on LS residual)}
\label{addLSdel_lscs}
LS-CS uses partial knowledge of support in a different way than modified-CS. It first computes an initial LS estimate using the known part of the support, $T$, and then computes the LS observation residual. CS is applied on the residual instead of applying it to the observation. Add-LS-del is used for support estimation.
%The LS-CS algorithm replaces the modified-CS step from Algorithm \ref{modcsalgo_2} by the CS-residual step.
We summarize the algorithm in Algorithm \ref{lscsalgo_2}.%The only difference from  Algorithm \ref{modcsalgo_2} is in the first step.

%is the same as the dynamic modified-CS algorithm but with step \ref{modcsstep} replaced by
%replaces step \ref{modcsstep} of the dynamic modified-CS algorithm by the following.%dynamic CS-residual (

\begin{algorithm}[h!]
\caption{{\bf \small LS-CS with Add-LS-Del}}
%At $t=0$, compute $\hat{x}_{0}$ as the solution of $\min_{\beta}  \|(\beta)\|_1 \ \text{s.t.} \ ||y_0 -A_0 \beta|| \le \eps$. Compute $\Nhat_0 = \{i \in [1,m] : |(\xhat_{0,modcs})_i| > \alpha \}$. \\
For $t\ge 0$, do
\ben
\item {\em Simple CS. } Do as in Algorithm \ref{modcsalgo_2}.

\item {\em CS-residual. }
\ben
\item Use $T:=\Nhat_{t-1}$ to compute the initial LS estimate, $\xhat_{t,\text{init}}$, and the LS residual, $\tty$, using
\label{initls}
\bea
(\xhat_{t,\text{init}})_{T} \se {A_{T}}^\dag y_t, \ \ (\xhat_{t,\text{init}})_{T^c} = 0 \nn \\
\tty \se  y_t - A \xhat_{t,\text{init}}  %y_t - A_{T} (\xhat_{t,\text{init}})_{T} %:= A \beta_t + w_t
\label{deftty0}
\eea

\item Do CS on the LS residual, i.e.  solve
\bea
\min_\beta \|\beta\|_1 \ s.t. \ \| \tty - A \beta \| \le \eps
\label{simplecs}
\eea
and denote its output by $\betahat_t$. Compute %$\xhat_{t,\CSres}$ using %(\ref{xhatcsres}).
\bea
\xhat_{t,\CSres} : = \betahat_t + \xhat_{t,\text{init}}.
\label{xhatcsres}
\eea
\een

\item {\em Additions / LS.} Compute $T_\dett$ and LS estimate using it as in  Algorithm \ref{modcsalgo_2}. Use $\xhat_{t,\CSres}$ instead of $\xhat_{t,modcs}$.
%\label{addls}
%\bea
%T_\dett \se T \cup \{i \in T^c: |(\xhat_{t,\CSres})_i| > \alpha_{add} \} \nn \\
%(\xhat_{t,\dett})_{T_\dett} \se {A_{T_\dett}}^\dag y_t, \ \ (\xhat_{t,\dett})_{T_\dett^c} = 0
%\eea

\item {\em Deletions / LS.} Compute $\tT$ and LS estimate using it as in Algorithm \ref{modcsalgo_2}.
%\label{delete}
%\bea
%\tT \se  T_{\dett} \setminus  \{i \in T_\dett: |(\xhat_{t,\dett})_i| \le \alpha_{del} \} \nn \\
%%\eea
%%Compute the LS estimate using $\tT$:
%%\bea
%(\xhat_{t})_{\tT} \se {A_{\tT}}^\dag  y_t, \ \ (\xhat_{t})_{\tT^c} = 0  %{A_{\Nhat_t}}^\ddag
%\label{finalls}
%\eea

\item Set $\Nhat_t = \tT$. Output $\xhat_t$. Feedback $\Nhat_t$.% Increment $t$. %Go to step \ref{step1}.

\een
%Feedback $\Nhat_t$, increment $t$, and go to step \ref{step1noiseless}.
\label{lscsalgo_2}
\end{algorithm}

The CS-residual step error can be bounded as follows. The proof follows in exactly the same way as that given in \cite{just_lscs} where CS is done using Dantzig selector instead of (\ref{simplecs}). We use (\ref{simplecs}) here to keep the comparison with modified-CS easier.
% Rewrite $\tty = A \beta_t + w_t$ where $\beta_t$ is a sparse-compressible signal. First bound the initial LS step error and then use Theorem 1.2 of \cite{candes_rip} to bound $\|\beta_t - \betahat_t\|$.
\begin{lemma}[CS-residual error bound \cite{just_lscs}]
If  $\|w\| \le \eps$, $\delta_{2|\Delta|} < (\sqrt{2}-1)/2$ and $\delta_{|T|} < 1/2$,
\bea
&& \|x - \xhat_{\CSres}\| \le C'(|T|,|\Delta|) \eps +  \theta_{|T|,|\Delta|} C''(|T|,|\Delta|) \|x_\Delta\| \nn \\
&& C'(|T|,|\Delta|)  \defn C_1(2|\Delta|) + \sqrt{2} C_2(2|\Delta|) \sqrt{\frac{|T|}{|\Delta|}} \nn \\
&& C''(|T|,|\Delta|) \defn 2 C_2(2|\Delta|) \sqrt{{|T|}}, \ \text{where} \nn \\
&& \text{$C_1(S)$ is defined in (\ref{defC1s}),~}  C_2(S) \defn 2\frac{1 + (\sqrt{2}-1) \delta_S}{1 - (\sqrt{2}+1) \delta_S}
\eea
\label{lscs_bnd}
\end{lemma}

\subsection{Stability result for LS-CS with Add-LS-Del}
The overall approach is similar to the one discussed in the previous section for modified-CS. The key difference is in the detection condition lemma, which we give below. Its proof is given in Appendix \ref{proof_detectcond_lscs}.

\begin{lemma}[Detection condition for LS-CS]
Assume that $\|w\| \le \eps$, $|T| \le S_T$ and $|\Delta| \le S_\Delta$.
Let $b:=\|x_\Delta\|_\infty$. For a $\gamma \le 1$, let $\Delta_1:=\{i \in \Delta: \gamma b \le |x_i| \le b \}$ and let $\Delta_2:= \Delta \setminus \Delta_1$. Assume that $|\Delta_1| \le S_{\Delta_1}$ and $\|x_{\Delta_2}\| \le \kappa b$. All $i \in \Delta_1$ will definitely get detected at the current time if $\delta_{2S_\Delta} < (\sqrt{2}-1)/2$, $\delta_{S_T} < 1/2$, $\theta_{S_T,S_\Delta} C''(S_T,S_\Delta) \le {\gamma}/{2(\sqrt{S_{\Delta_1}} + \kappa)}$, and
\bea
%&&   \theta_{S_T,S_\Delta} \sqrt{S_{\Delta_1}} C''(S_T,S_\Delta) < {\gamma}/2 \ \ \text{and} \nn \\  %\max_{|\Delta| \le S_\Delta}
&& \max_{|\Delta| \le S_\Delta} \frac{ \alpha_{add} + C'(S_T,|\Delta|)\eps }{\gamma - {\theta_{S_T,|\Delta|}} C''(S_T,|\Delta|) (\sqrt{S_{\Delta_1}} + \kappa)} < b \ \  \nn
\label{detcond_lscs}
\eea
where $C'(.,.)$, $C''(.,.)$ are defined in Lemma \ref{lscs_bnd}.
\label{detectcond_lscs}
\end{lemma}

The stability result then follows in the same fashion as Theorem \ref{stabres_modcs}. The only difference is that instead of Lemma \ref{detectcond_modcs}, we apply Lemma \ref{detectcond_lscs} with $S_T=S_0, S_\Delta=2S_a$, $b=2r$, $\gamma=1$, $S_{\Delta_1}=S_a$ and $\kappa = (\sqrt{2S_a} r)/(2r) = \sqrt{S_a}/\sqrt{2}$.

\begin{theorem}[Stability of LS-CS]
Assume Signal Model \ref{sigmod2} and $\|w\| \le \eps$. If %the following hold for some $1 \le d_0 \le d$,%bounded noise with
\ben
\item  {\em (addition and deletion thresholds) }
\ben
\item $\alpha_{add}$ is large enough so that there are at most $S_a$ false additions per unit time,
\label{addthresh_lscs}

%\item $\alpha_{del}  = \sqrt{\frac{2}{S_0-S_a}} \eps +  \sqrt{\frac{8S_a}{S_0-S_a}} \theta_{S_0+2S_a,S_a} r $,
\item $\alpha_{del}  = \sqrt{2} \eps + 2 \sqrt{S_a} \theta_{S_0+2S_a,S_a} r $
\label{delthresh_lscs}
\een

\item {\em (support size, support change rate)} $S_0, S_a$ satisfy
\label{measmodel_lscs}
\ben
\item $\delta_{4S_a} < (\sqrt{2}-1)/2$
\label{measmodel_lscs_1a}
\item $\delta_{S_0+2S_a} < 1/2$%$k^a S_a \le \sinf$, $S_0+S_a + f \le \sone$,
\label{measmodel_lscs_1b}

\item $\theta_{S_0,2S_a} C''(S_0,2S_a)   <  \frac{1}{(2+\sqrt{2})\sqrt{S_a}}$ %\max_{|\Delta| \le k_1(d_0)}
\label{measmodel_lscs_2a}

\item $\theta_{S_0+2S_a,S_a} < \frac{1}{2} \sqrt{  \frac{1}{4S_a } }$
\label{measmodel_lscs_2b}
%\item $\theta_{S_0+2S_a,S_a} \le  \frac{1}{2} \sqrt{ \frac{S_0-S_a}{8S_a} }$

\een

\item {\em  (new element increase rate) } $ r \ge \max(\tilde{G}_1,\tilde{G}_2)$, where
\label{add_del_lscs}
\bea
&& \tilde{G}_1 \defn
 \max_{|\Delta| \le 2S_a} %\nn \\ &&
 [\frac{ \alpha_{add} + C'(S_0,|\Delta|)\eps  }{2 -  (2+\sqrt{2}) {\theta_{S_0,|\Delta|}} \sqrt{S_a} C''(S_0,|\Delta|)}] \nn \\
%
%&& \tilde{G}_2 \defn \frac{\sqrt{2} \eps}{\sqrt{S_0-S_a} (1 - \theta_{S_0+2S_a,S_a}\sqrt{ \frac{8S_a}{S_0-S_a}})}
&& \tilde{G}_2 \defn \frac{\sqrt{2} \eps}{1 - \theta_{S_0+2S_a,S_a}\sqrt{4S_a }}
\eea
%\alpha_{del} + \sqrt{2} \eps + 2{\theta{(S_0+S_a+f,k_2(d_0))}} \sm(d_0) \nn

\item {\em (initialization) }  (same condition as in Theorem \ref{stabres_modcs})
\een
then, all conclusions of Theorem \ref{stabres_modcs} hold for LS-CS, except the last one, which is replaced by $\|x_t - \xhat_{t,\CSres}\| \le \max_{|\Delta| \le 2S_a} [ C'(S_0,|\Delta|)\eps  + (\theta_{S_0,|\Delta|} C''(S_0,|\Delta|) + 1) \sqrt{2S_a} r]$.%[2S_a r^2 + 4S_a r^2]]$.%Also, all conclusions of Corollary \ref{stabres_modcs_cor} hold except the last one.dynamic CS-residual (
\label{stabres_lscs}
\end{theorem}

\subsection{Discussion}
Notice that conditions \ref{measmodel_lscs_2a} and \ref{measmodel_lscs_2b} are the most difficult conditions to satisfy as the problem size increases and consequently $S_a$ increases. We get condition \ref{measmodel_lscs_2b} because we bound the $\ell_\infty$ norm of the detection LS step error by its  $\ell_2$ norm which is a loose bound. This can be relaxed in the same fashion as in the previous section by assuming that the LS step error is spread out enough (see Corollary \ref{cor2_relax}).

\begin{figure*}[t!]
\centerline{
\subfigure[$r=3/3, d=3, M=dr=3$]{
\label{highsnr}
\begin{tabular}{ccc}
\epsfig{file = 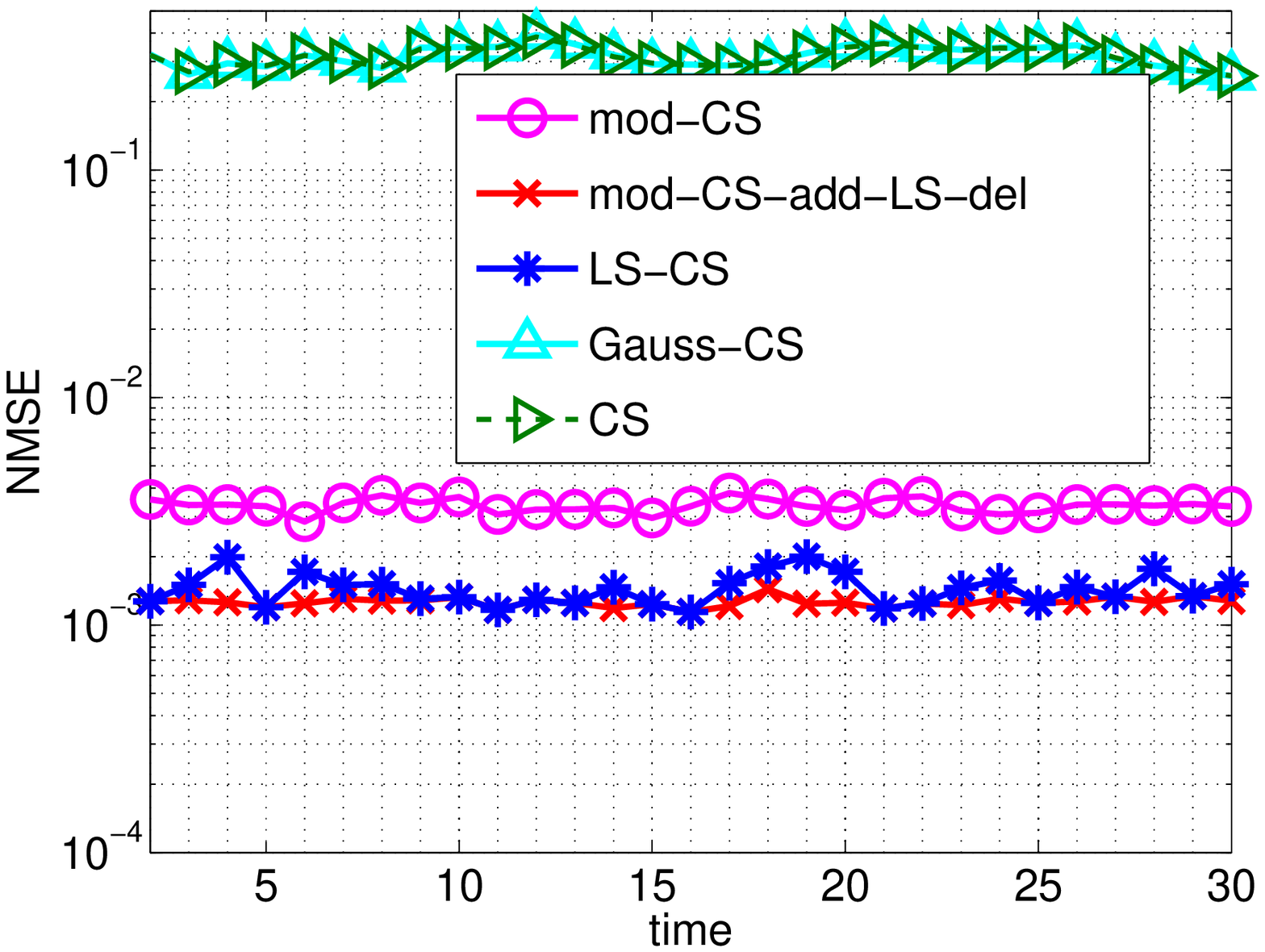, width=4cm, } &
\epsfig{file = 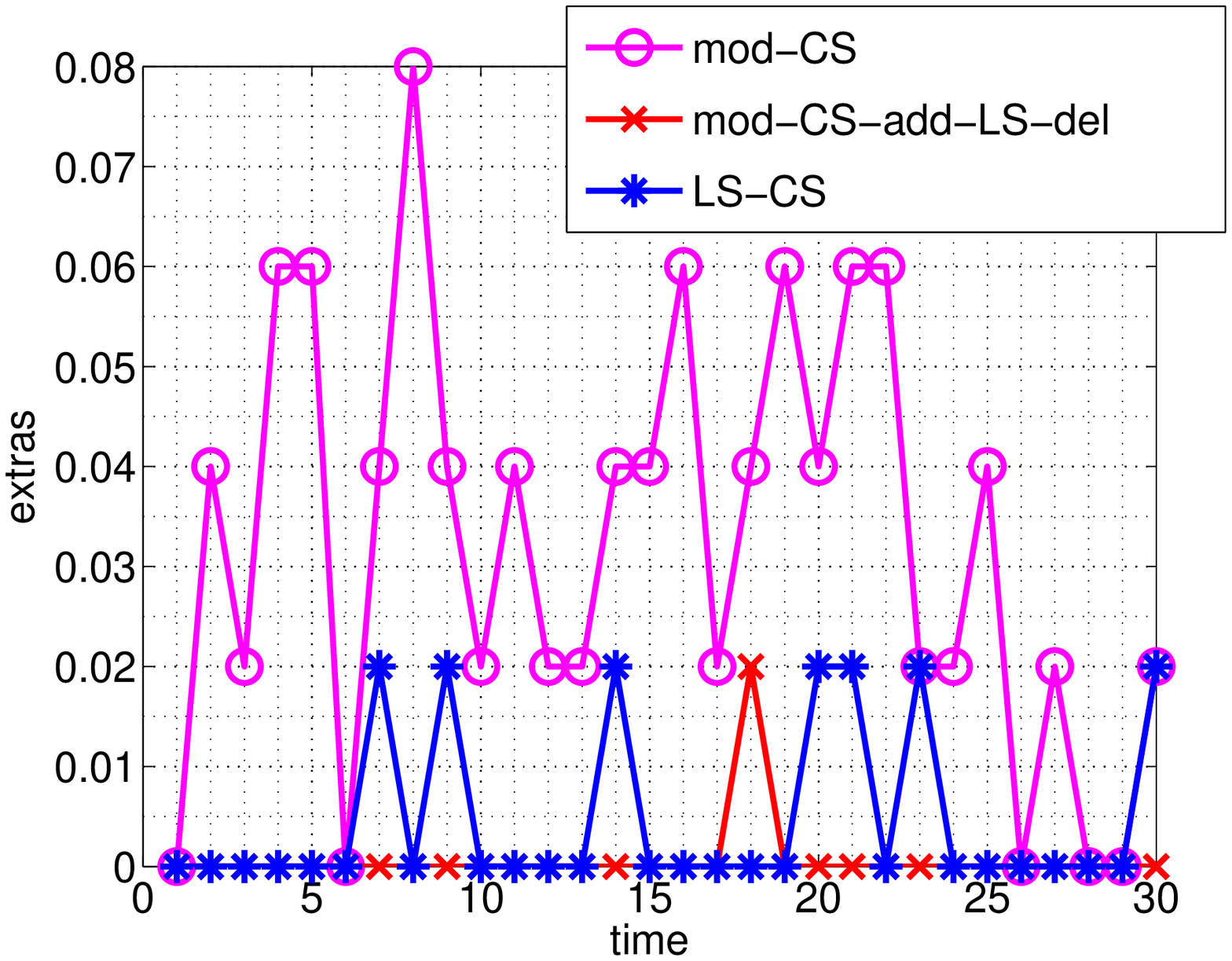, width=4cm} &
\epsfig{file = 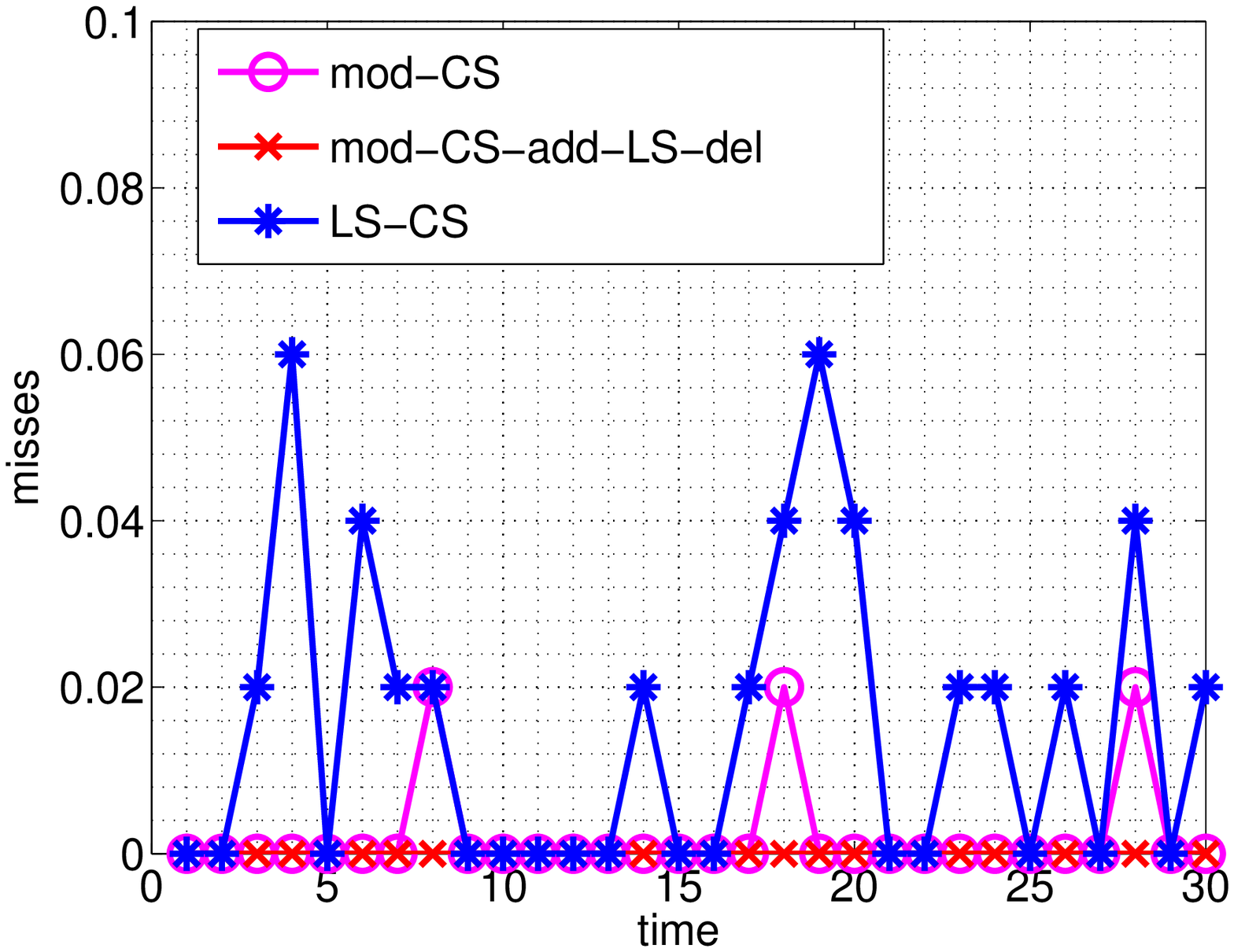, width=4cm}
\end{tabular}
}
}
\centerline{
\subfigure[$r=3/4, d=4, M=dr=3$]{
\label{r3_4}
\begin{tabular}{ccc}
\epsfig{file = 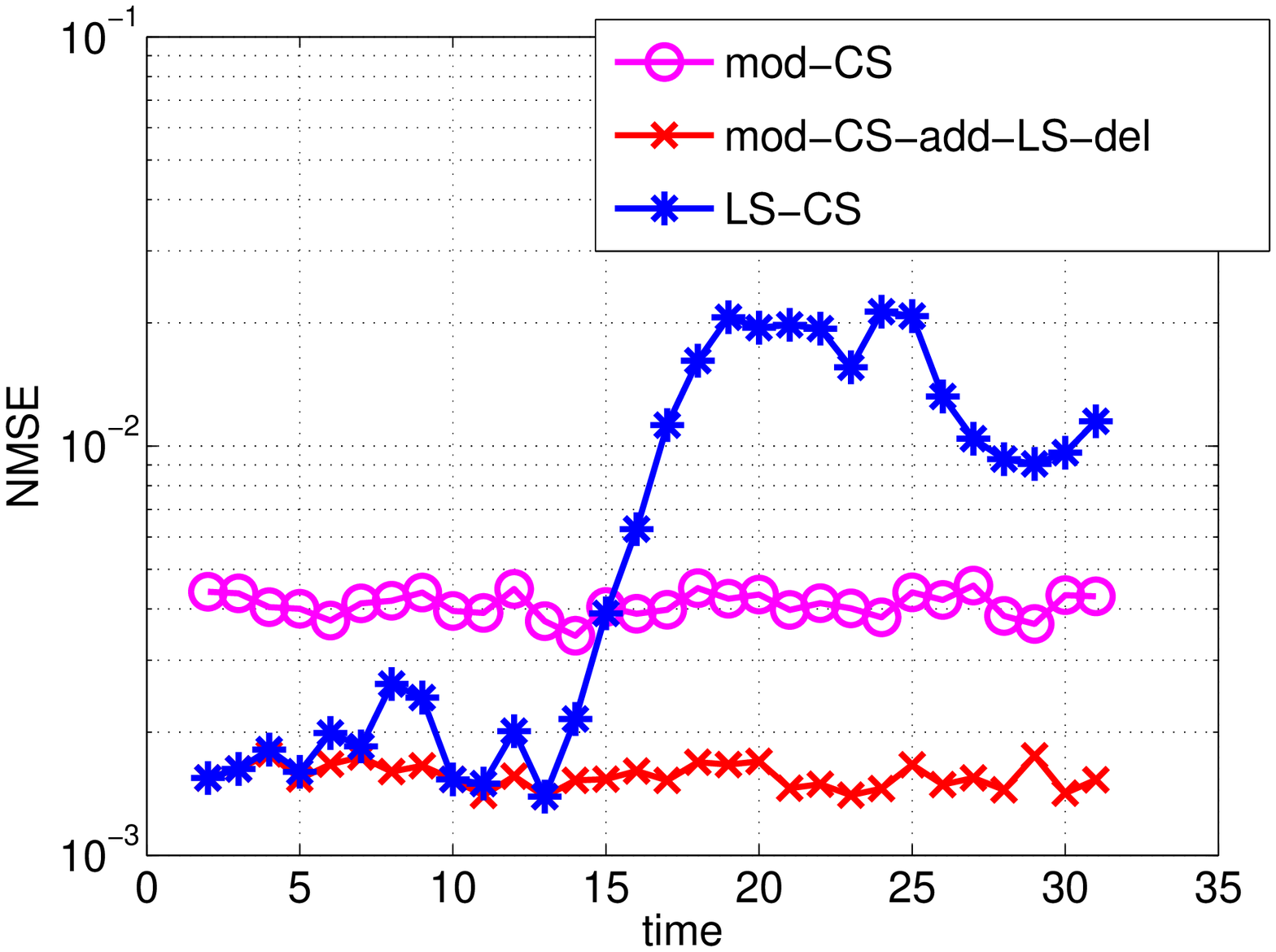, width=4cm} &
\epsfig{file = 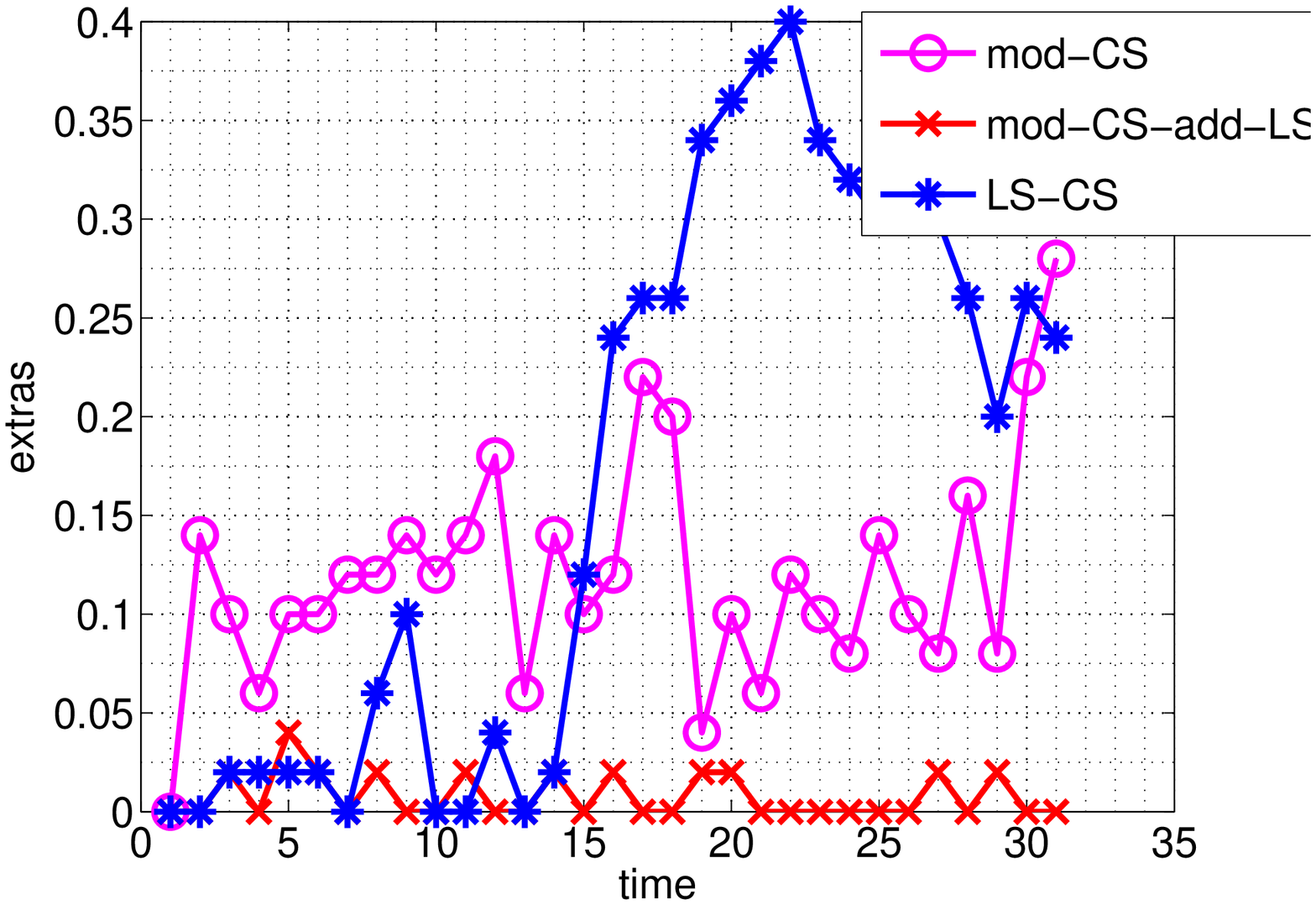, width=4cm} &
\epsfig{file = 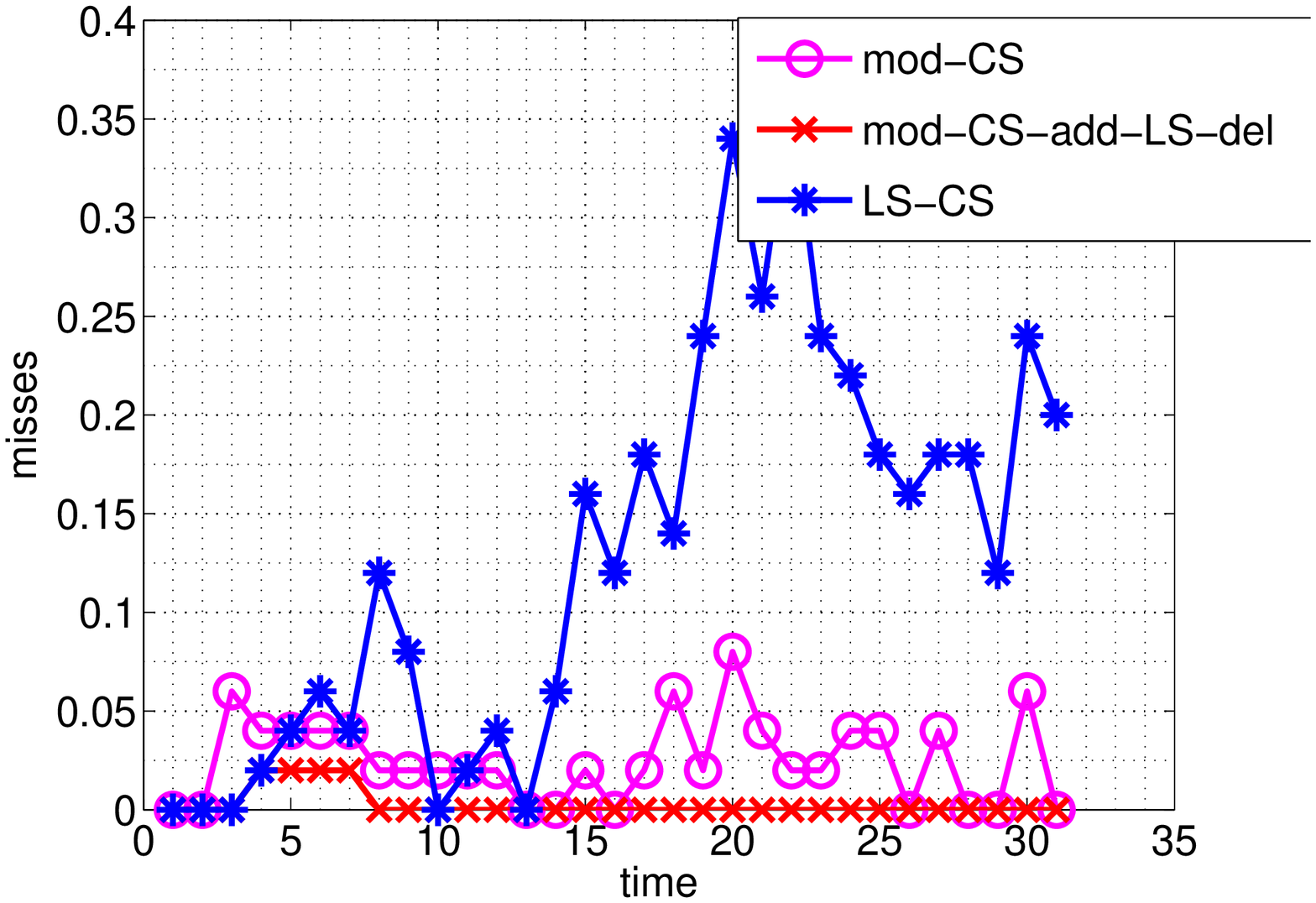, width=4cm}
\end{tabular}
}
}
\centerline{
\subfigure[$r=2/4, d=4, M=dr=2$]{
\label{r2_4}
\begin{tabular}{ccc}
\epsfig{file = 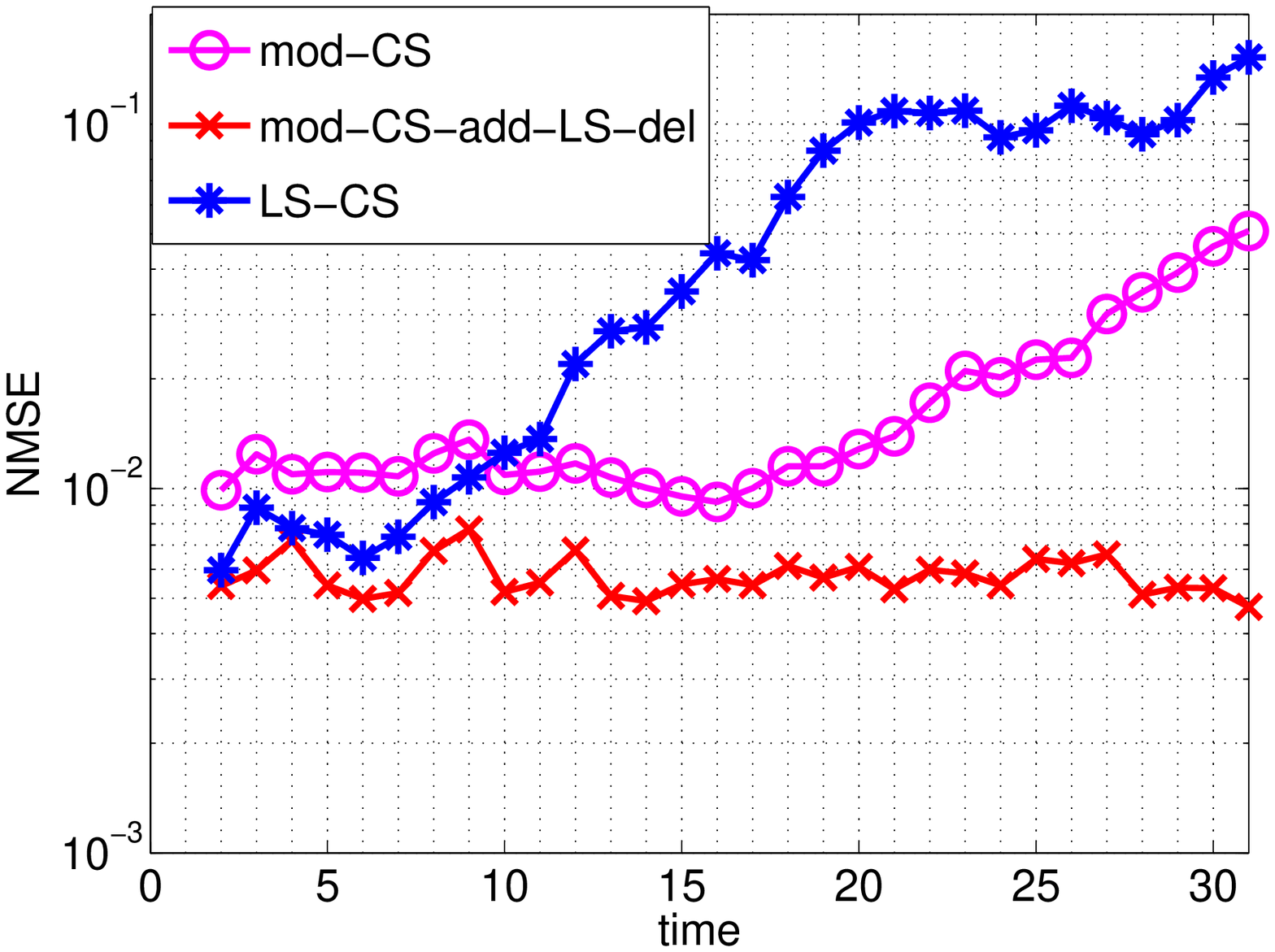, width=4cm} &
\epsfig{file = 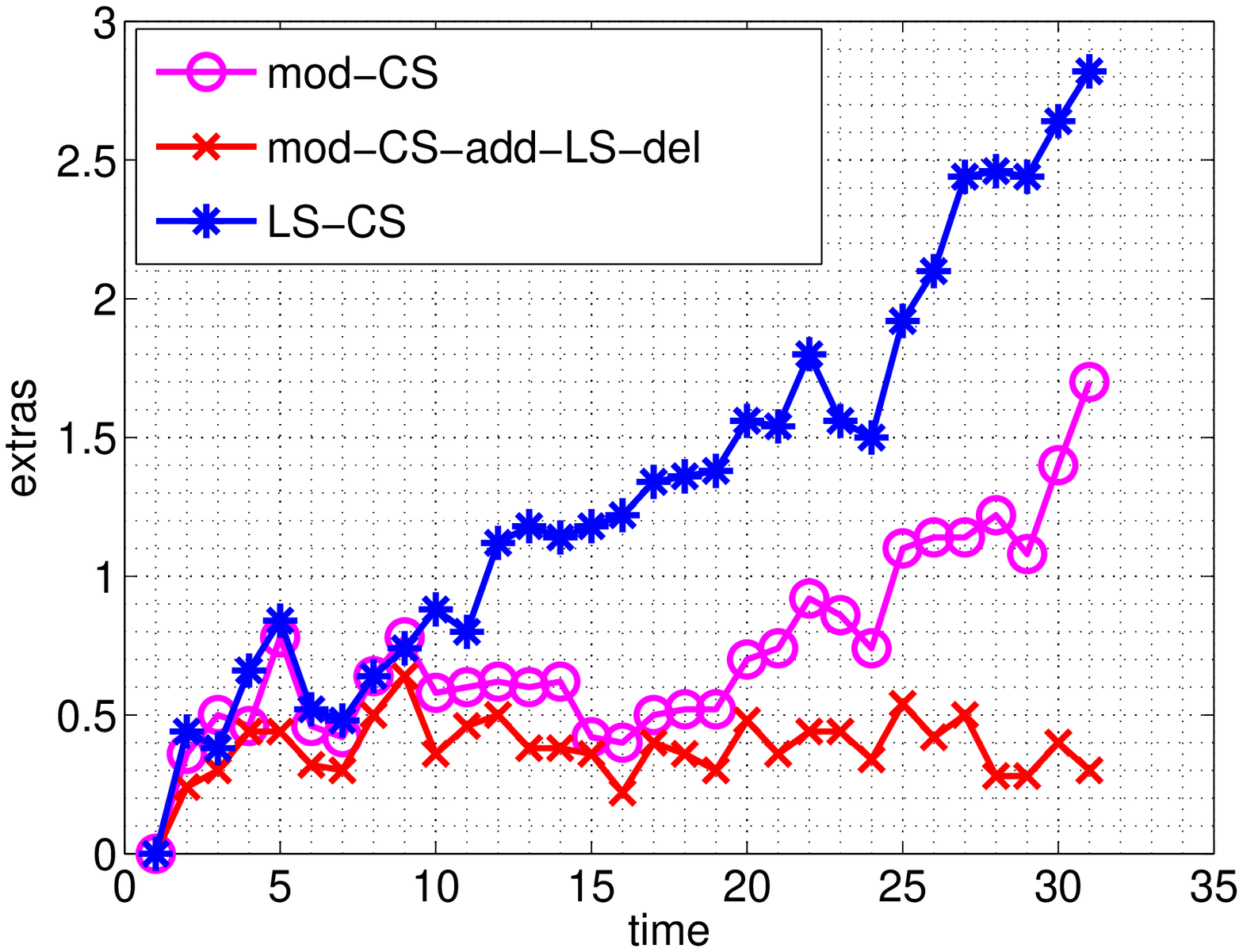, width=4cm} &
\epsfig{file = 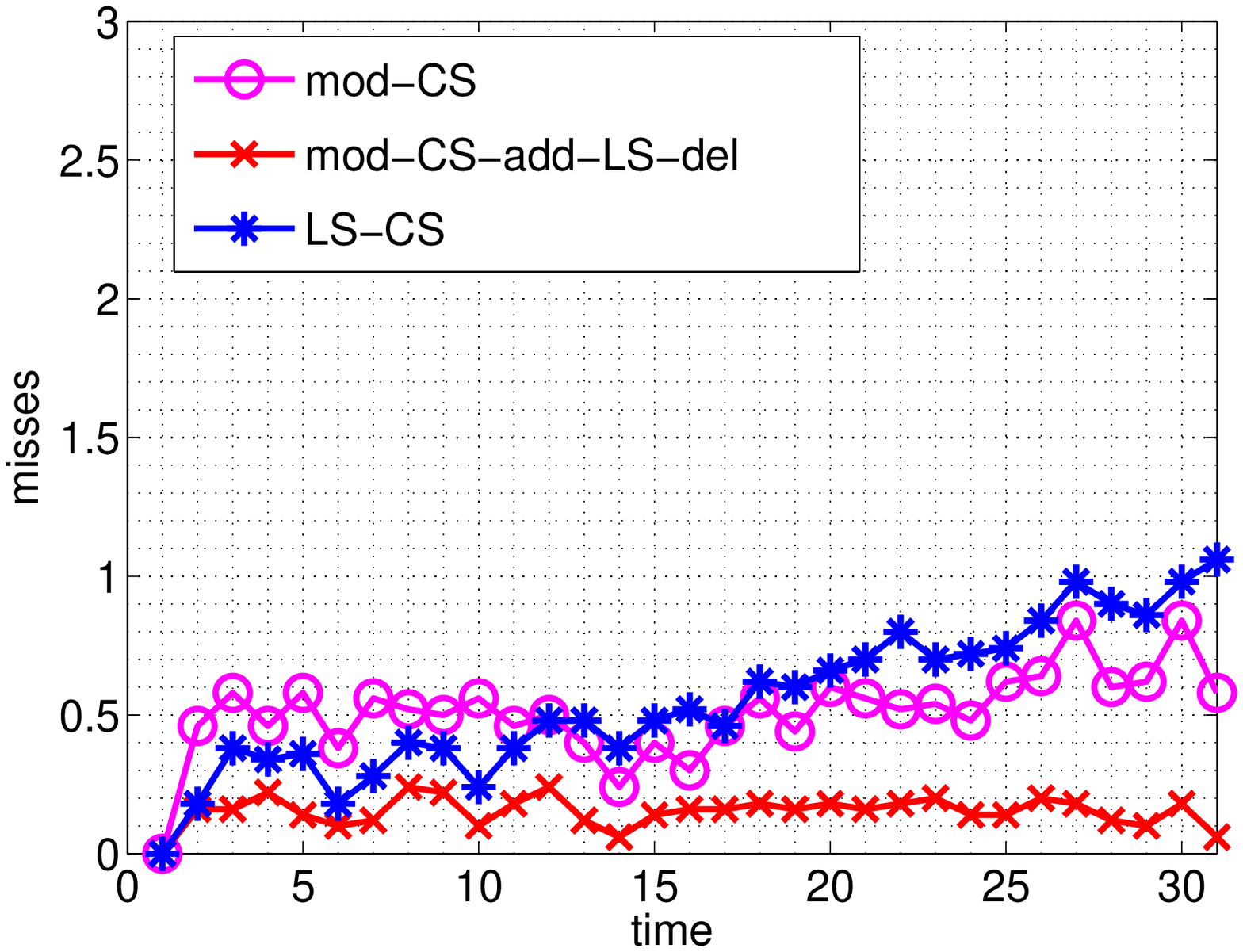, width=4cm}
\end{tabular}
}
}
\centerline{
\subfigure[$r=2/5, d=5, M=dr=2$]{
\label{r2_5}
\begin{tabular}{ccc}
\epsfig{file = 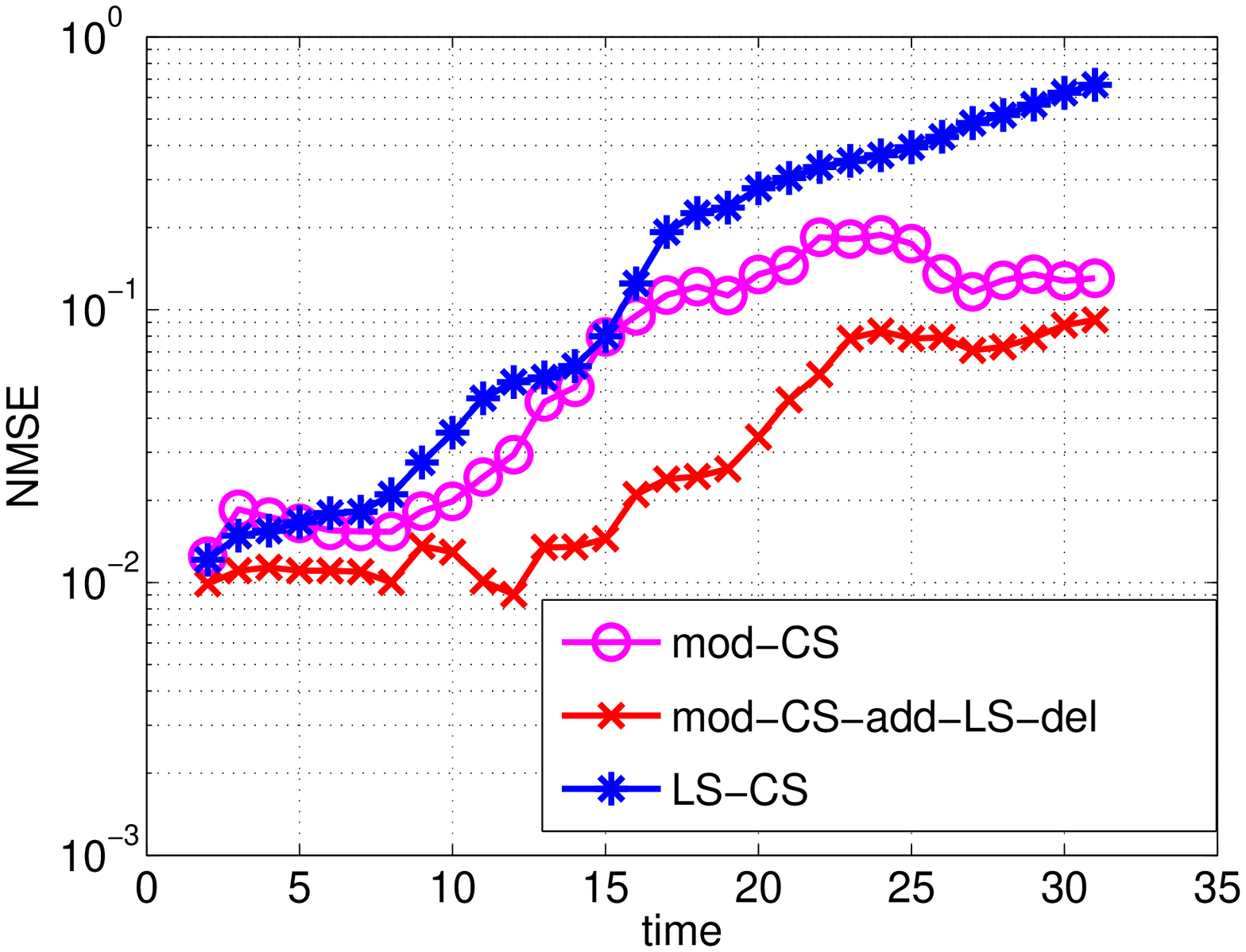, width=4cm} &
\epsfig{file = 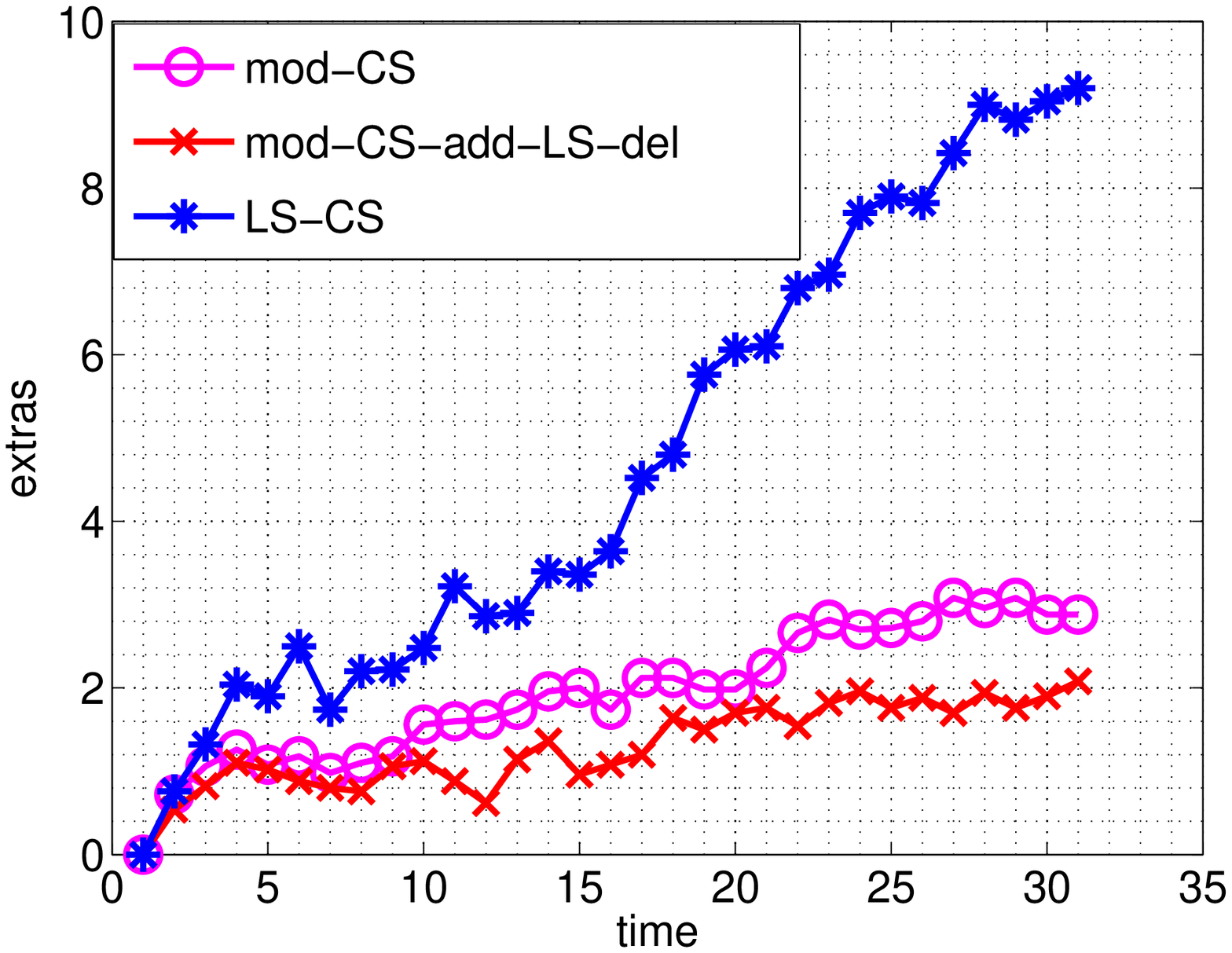, width=4cm} &
\epsfig{file = 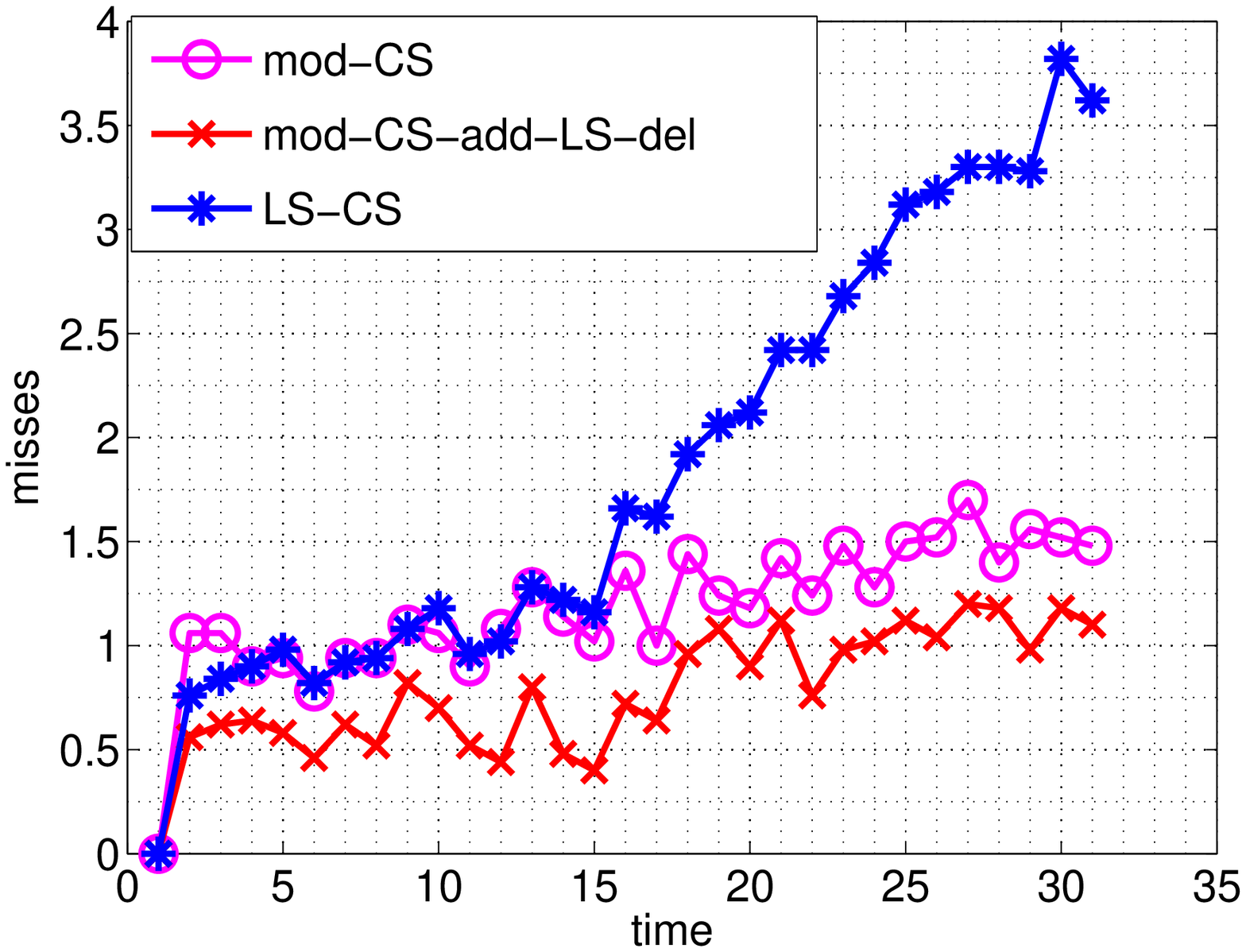, width=4cm}
\end{tabular}
}
}
%\centerline{
%\subfigure[$r=3/5, d=5, M=dr=3$]{
%\begin{tabular}{ccc}
%\epsfig{file = C:/Code/KFCS/LSCS_KFCS_code/nmse_M3_d5_n59.eps, width=4cm} &
%\epsfig{file = C:/Code/KFCS/LSCS_KFCS_code/extras_M3_d5_n59.eps, width=4cm} &
%\epsfig{file = C:/Code/KFCS/LSCS_KFCS_code/misses_M3_d5_n59.eps, width=4cm}
%\end{tabular}
%}
%}
\vspace{-0.1in}
\caption{\small{Normalized MSE (NMSE), number of extras and number of misses over time for modified-CS, modified-CS with add-LS-del and LS-CS. In all cases, NMSE for CS was more than 20\% in all cases (plotted only in (a)).
}}
\vspace{-0.15in}
\label{fig1}
\end{figure*}

Consider condition \ref{measmodel_lscs_2a}. We get this because (i) we upper bound the $\ell_\infty$ norm of the CS-residual step error, $x_t - \xhat_{t,CSres}$, by its $\ell_2$ norm and (ii) in the proof of Lemma \ref{lscs_bnd}, we upper bound the $\ell_1$ norm of the initial LS step error, $(x_t - \xhat_{t,\text{init}})_T$, by $\sqrt{|T|}$ times its $\ell_2$ norm (this results in the expression for $C''(.,.)$ given in Lemma \ref{lscs_bnd}). If one can argue that both the initial LS step error and the CS-residual error are spread out enough, we can relax condition \ref{measmodel_lscs_2a} to make it somewhat comparable to that of modified-CS. %then get rid of the $\sqrt{S_a}$ and $\sqrt{S_0}$ terms in condition \ref{measmodel_lscs_2a}.  with add-LS-del result
%The LS-CS result will then become somewhat comparable to the modified-CS stability result.
But even then, $\tilde{G}_1$ will be larger and so LS-CS will still require a higher rate of coefficient increase, $r$, to ensure stability. This is also observed in our simulations. See Fig. \ref{r2_4}.%Notice that $C''(S_0,2S_a) = 2C_1(4S_a) \sqrt{S_0}$.

\section{Simulation Results}
\label{sims}
We compared modified-CS (mod-cs), modified-CS with Add-LS-Del (mod-cs-add-del), LS-CS and simple CS for a few different choices of $r$. In all cases, we used Signal Model \ref{sigmod2} with $m=200$, $S_0=20$, $S_a=2$, $n=59$ and $w_t \sim^{i.i.d.} uniform(-c,c)$ with $c=0.1266$. The specific generative model that we used is specified in Appendix \ref{generativemodel} and also briefly discussed in Sec. \ref{signalmodel}.
The measurement matrix was random-Gaussian. We averaged over 50 simulations. In all cases, we set the addition threshold, $\alpha_{add}$, to be at the noise level - we set it to $c/2$.
Assuming that the LS step after addition gives a fairly accurate estimate of the nonzero values, one can set the deletion threshold, $\alpha_{del}$, to a larger value of $r/2$ and still ensure that there are no (or very few) false deletions. Larger deletion threshold ensures that all (or most) of the false additions and removals get deleted. Modified-CS used a single threshold, $\alpha$, somewhere in between $\alpha_{add}$ and $\alpha_{del}$. We set $\alpha = ((c/2)+(r/2))/2$.
%used $\alpha_{add} = c/2 = 0.0633$, $\alpha_{del} = r/2$ (used by modified-CS-add-del and LS-CS) and $\alpha = ((c/2)+(r/2))/2$ (used by modified-CS).

In Fig. \ref{fig1}, we show two sets of plots: $r=1$, $d=3$ ($M=dr=3$) in \ref{highsnr} and  $r=0.5$, $d=4$ ($M=dr=2$) in  \ref{r2_4}.
%four sets of plots: $r=1$, $d=3$ ($M=dr=3$) in \ref{highsnr}; $r=0.75$, $d=4$ ($M=dr=3$) in \ref{r3_4} and  $r=0.5$, $d=4$ ($M=dr=2$) in  \ref{r2_4} and $r=0.4$, $d=5$ ($M=dr=2$) in \ref{r2_5}.  and $d$
Normalized MSE (NMSE), average number of extras (mean of $|\Nhat_t \setminus N_t|$ over the 50 simulations) and average number of misses (mean of $|N_t \setminus \Nhat_t|$) are plotted in the left, middle and right columns respectively. Notice that when the support size is $S_0=20$, $n=59$ is too small for CS to work and hence in all cases, the NMSE of CS was more than 20\%. We show its NMSE only in \ref{highsnr}.

When $r=1$, all of mod-cs, mod-cs-add-del and ls-cs are stable. Mod-cs-add-del uses a better support estimation method (add-LS-del) and thus its extras and misses are both much smaller than those of mod-cs (in this case, it is possible that if we experimented with many different threshold choices, mod-cs error could be made smaller). %\footnote{Of course, it is possible that if we tried many different threshold choices, mod-cs error could also become smaller}.
As a result its reconstruction error is also stable at a smaller value. In this case, since $r$ is large enough, LS-CS (which also uses add-LS-del) has similar error to that of Mod-cs-add-del.
%
%Notice that since mod-cs uses a single threshold for addition/deletion, which in this case happens to be too large, so that the number of extras is larger than that of either of the other methods. Hence it has the largest NMSE.
%Its number of misses is also larger than mod-cs-add-del (which has the flexibility of picking two thresholds).
%
When $r$ is reduced to 0.75, it becomes too small for LS-CS and so LS-CS becomes unstable. LS-CS stability is discussed in Sec. \ref{addLSdel_lscs}. As we explain there, due to the CS-residual step, LS-CS needs a larger $r$ for stability.
When $r$ is reduced to 0.5, mod-CS also becomes unstable. But mod-cs-add-del is still stable.
 Mod-CS uses one threshold and hence as explained after Theorem \ref{stabres_simple_modcs}, it needs a larger $r$ for stability than mod-cs-add-del.
Finally if we reduce $r$ to 0.4, all three became unstable.%(not shown, shown in the long version \cite{long}).  (not shown here but see Fig. 1 of \cite{long})

\section{Conclusions}
\label{conclusions}
We showed the ``stability" of modified-CS and its improved version, modified-CS with add-LS-Del, and of LS-CS for signal sequence reconstruction, under mild assumptions. By ``stability" we mean that the number of misses from the current support estimate and the number of extras in it remain bounded by a time-invariant value at all times. The result is meaningful when the bound is small compared to the support size.%A direct corollary is a time-invariant and small bound on the
%and the value is small compared to the signal support size.% A direct corollary of this would be that the reconstruction errors are also small compared to the signal power.. The concept is meaningful only if the bound

\appendix
\subsection{A generative model for  Signal Model \ref{sigmod2}}
\label{generativemodel}
To help understand the model better (and also to simulate it), we describe here one plausible generative model that satisfies its required assumptions. This assumes that every element gets added to the support at magnitude $r$ and keeps increasing until it reaches magnitude $M$. Similarly, every element that began decreasing keeps decreasing until it becomes zero and gets removed from the support\footnote{Another possible generative model is: select $S_a$ out of the $2S_a$ current elements with magnitude $jr$ and increase them, and decrease the other $S_a$ elements}. We keep the signs of the elements the same except when the element first gets added (at that time, can set the sign to $\pm 1$ with equal probability).%
%Also, we keep the signs the same except when the coefficient first gets added (at that time, can set the sign to $\pm 1$ with equal probability).
%A generative model for Signal Model \ref{sigmod2} can be specified in the following manner. First define

To specify the generative model, first define
\bd
Define
\ben
\item Increasing set, $\Inc_t := \Iset_t(1) \cup \Iset_t(2) \dots \cup  \Iset_t(d-1)$
\item Decreasing set, $\Dec_t := \Dset_t(1) \cup \Dset_t(2) \dots \cup  \Dset_t(d-1)$
\item Constant set, $\Con_t:= \{i : |x_{t,i}| = M \}$. Clearly $\Con_t = N_t \setminus (\Inc_t \cup \Dec_t)$.
\een
\ed

%Then at any time $t > 0$, $N_t = \Inc_t \cup \Dec_t \cup \Con_t$ and $|N_t|=S_0$. The set $\Inc_t$ contains $S_a$ elements each with magnitude $r, 2r, \dots (d-1)r$ and similarly for $\Dset_t$. The set $\Con_t$ contains $S_0-(2d-2)S_a$ elements with magnitude $M$.

The generative model is as follows. At each $t>0$,
\ben
%\item At time $t$, select the set of coefficients which begin to decrease, i.e. select $\Dset_t(d-1) \subseteq \Con_{t-1}$ of size $S_a$ and select the newly added set, $\Iset_t(1) \subseteq N_{t-1}^c$ of size $S_a$.

\item Update the magnitudes for elements of the previous increasing, decreasing and constant sets.
\bea
(x_{t})_{\Inc_{t-1}}  \se [ |(x_{t-1})_{\Inc_{t-1}}| + r ] \ \sgn((x_{t-1})_{\Inc_{t-1}})   \nn \\
(x_{t})_{\Dec_{t-1}} \se [|(x_{t-1})_{\Dec_{t-1}}| - r] \ \sgn((x_{t-1})_{\Dec_{t-1}}) \nn \\
(x_{t})_{\Con_{t-1}} \se (x_{t-1})_{\Con_{t-1}}
\eea
where $\sgn(z)$ is a vector containing the signs of each element of $z$.

\item Select the newly added set, $\Iset_t(1) \subseteq N_{t-1}^c$, of size $S_a$ uniformly at random. Similarly select the new set of decreasing elements, $\Dset_t(d-1) \subseteq \Con_{t-1}$ of size $S_a$ uniformly at random. Set their values as:
\bea
(x_{t})_{\Iset_t(1)} \se r \ \underline{s}    \nn \\
(x_{t})_{\Dset_t(d-1)} \se (d-1)r \ \sgn((x_{t-1})_{\Dset_{d-1}})
\eea
where $\underline{s}$ is an $S_a \times 1$ signs' vector in which each element is $+1$ or $-1$ with probability $1/2$.

\item Compute:
\bea
\Iset_t(d) \se  \{i \in \Inc_{t-1} : |(x_{t-1})_i| = (d-1)r \} \nn \\
\Dset_t(0) \se  \{i \in \Dec_{t-1} : |(x_{t-1})_i| = r \}
\eea
and update the increasing, decreasing and constant sets:
\bea
%\Iset_t(j) \se \Iset_{t-1}(j-1), \ \forall \ j=2, \dots (d-1) \nn \\ %\Iset_t(1) \se \Aset_t, \
%\Dset_t(j) \se \Dset_{t-1}(j+1), \ \forall \ j=1, \dots (d-2) \nn \\
\Inc_t \se \Inc_{t-1} \cup \Iset_t(1) \setminus \Iset_t(d) \nn \\
\Dec_t \se \Dec_{t-1} \cup \Dset_t(d-1) \setminus \Dset_t(0) \nn \\
\Con_t \se \Con_{t-1} \cup  \Iset_t(d) \setminus \Dset_t(d-1) \nn \\
N_t \se \Inc_t \cup \Dec_t \cup \Con_t
\eea

%\item  The model does not assume anything about signs. But a reasonable generative model will keep the signs the same except when the coefficient first gets added (at that time, can set the sign to $\pm 1$ with equal probability).
\een

\subsection{Appendix: Proof of Theorem \ref{stabres_simple_modcs}}% and Corollary \ref{stabres_modcs_cor}
\label{proof_simple_modcs}

We prove the first claim by induction. Using condition \ref{initass_simple} of the theorem, the claim holds for $t=0$. This proves the base case. For the induction step, assume that the claim holds at $t-1$, i.e. $|\tDelta_{e,t-1}| =0$, $|\tT_{t-1}| \le S_0$, and $\tDelta_{t-1} \subseteq \Sset_{t-1}(2)$ so that $|\tDelta_{t-1}| \le 2S_a$. Using this assumption we prove that the claim holds at $t$. In the proof, we use the following facts often: (a) $\Rset_t \subseteq N_{t-1}$ and $\Aset_t \subseteq N_{t-1}^c$, (b) $N_t = N_{t-1} \cup \Aset_t \setminus \Rset_t$, and (c) if two sets $B,C$ are disjoint, then, $(D \cap B^c) \cup C = D \cup C \setminus B$ for any set $D$.%

We first bound $|T_t|$, $|\Delta_{e,t}|$ and $|\Delta_t|$.
Since $T_t = \tT_{t-1}$, so $|T_t| \le S_0$. Since $\Delta_{e,t} = \Nhat_{t-1} \setminus N_t =  \Nhat_{t-1} \cap [(N_{t-1}^c \cap \Aset_t^c) \cup \Rset_t] \subseteq \tDelta_{e,t-1} \cup \Rset_t = \Rset_t$. The last equality follows since $|\tDelta_{e,t-1}| =0$. Thus $|\Delta_{e,t}| \le |\Rset_t| = S_a$.
Now consider $|\Delta_t|$. Notice that $\Delta_t = N_t \setminus \Nhat_{t-1}  = (N_{t-1} \cap \Nhat_{t-1}^c \cap \Rset_t^c) \cup (\Aset_t \cap \Rset_t^c \cap \Nhat_{t-1}^c) = (\tDelta_{t-1} \cap  \Rset_t^c) \cup ( \Aset_t \cap \Nhat_{t-1}^c) \subseteq  (\Sset_{t-1}(2)  \cap  \Rset_t^c) \cup  \Aset_t = \Sset_{t-1}(2) \cup  \Aset_t \setminus \Rset_t$. The second last equality uses $\tDelta_{t-1} \subseteq \Sset_{t-1}(2)$.
 Since $\Rset_t$ is a subset of $\Sset_{t-1}(2)$ and $\Aset_t$ is disjoint with $\Sset_{t-1}(2)$, thus $|\Delta_t| \le |\Sset_{t-1}(2)| + |\Aset_t| - |\Rset_t| = 2S_a + S_a - S_a = 2S_a$.
% %The last one follows from the disjointness of $\Rset_t$ and $\Aset_t$.
%These follow from the definition of $\tDelta_t$, disjointness of $\Rset_t$ and $\Aset_t$ and using $\tDelta_{t-1} \subseteq \Sset_{t-1}(2)$ (induction assumption). %The last and third last one follow since $\Rset_t$ and $\Aset_t$ are disjoint.%(\tDelta_{t-1} \cap  \Rset_t^c) \cup  \Aset_t \subseteq

Next we bound  $|\tDelta_t|$, $|\tDelta_{e,t}|$ and $|\tT_t|$. Consider the support estimation step. Apply the first claim of Lemma \ref{lemma_modcs} with $S_N=S_0$, $S_{\Delta_e}=S_a$, $S_\Delta = 2S_a$, and $b_1 = 2r$. Since conditions \ref{measmodel_simple} and \ref{add_del_simple} of the theorem hold, all elements of $N_t$ with magnitude equal or greater than $2r$ will get detected. Thus, $\tDelta_t \subseteq \Sset_t(2)$. Apply the second claim of the lemma. Since conditions \ref{measmodel_simple} and \ref{threshes_simple} hold, all zero elements will get deleted and there will be no false detections, i.e. $|\tDelta_{e,t}|=0$. Finally using $|\tT_t| \le |N_t| + |\tDelta_{e,t}|$, $|\tT_t| \le S_0$.

\subsection{Appendix: Proof of Theorem \ref{stabres_modcs}}% and Corollary \ref{stabres_modcs_cor}
\label{proof_addLSdel_modcs}

We prove the first claim of the theorem by induction. Using condition \ref{initass} of the theorem, the claim holds for $t=0$. This proves the base case. For the induction step, assume that the claim holds at $t-1$, i.e. $|\tDelta_{e,t-1}| =0$, $|T_{t-1}| \le S_0$, and $\tDelta_{t-1} \subseteq \Sset_{t-1}(2)$ so that $|\tDelta_{t-1}| \le 2S_a$. Using the induction assumption, we prove that the claim holds at $t$. In the proof, we will use  the following facts often:  $\Rset_t \subseteq N_{t-1}$, $\Aset_t \subseteq N_{t-1}^c$ and  $N_t = N_{t-1} \cup \Aset_t \setminus \Rset_t$. Also, if two sets $B,C$ are disjoint, then, $(D \cap B^c) \cup C = D \cup C \setminus B$ for any set $D$.%equation (\ref{sseteq}) and

Since $T_t = T_{t-1}$, so $|T_t| \le S_0$.
Since $\Delta_{e,t} = \Nhat_{t-1} \setminus N_t =  \Nhat_{t-1} \cap [(N_{t-1}^c \cap \Aset_t^c) \cup \Rset_t] \subseteq \tDelta_{e,t-1} \cup \Rset_t = \Rset_t$. The last equality follows since $|\tDelta_{e,t-1}| =0$. Thus $|\Delta_{e,t}| \le |\Rset_t| = S_a$.
Next we bound $|\Delta_{t}|$. Note that $\Delta_t = N_t \setminus \Nhat_{t-1}  = (N_{t-1} \cap \Nhat_{t-1}^c \cap \Rset_t^c) \cup (\Aset_t \cap \Rset_t^c \cap \Nhat_{t-1}^c) = (\tDelta_{t-1} \cap  \Rset_t^c) \cup ( \Aset_t \cap \Nhat_{t-1}^c) \subseteq  (\Sset_{t-1}(2)  \cap  \Rset_t^c) \cup  \Aset_t = \Sset_{t-1}(2) \cup  \Aset_t \setminus \Rset_t$. Since $\Rset_t$ is a subset of $\Sset_{t-1}(2)$ and $\Aset_t$ is disjoint with $\Sset_{t-1}(2)$, thus $|\Delta_t| \le |\Sset_{t-1}(2)| + |\Aset_t| - |\Rset_t| = 2S_a + S_a - S_a$.
 %The last one follows from the disjointness of $\Rset_t$ and $\Aset_t$.
%These follow from the definition of $\tDelta_t$, disjointness of $\Rset_t$ and $\Aset_t$ and using $\tDelta_{t-1} \subseteq \Sset_{t-1}(2)$ (induction assumption). %The last and third last one follow since $\Rset_t$ and $\Aset_t$ are disjoint.%(\tDelta_{t-1} \cap  \Rset_t^c) \cup  \Aset_t \subseteq

Consider the detection step. There are at most $S_a$ false detects (from condition \ref{addthresh}) and thus $|\tDelta_{e,\dett,t}| \le |\Delta_{e,t}| + S_a \le 2S_a$. Thus $|T_{\dett,t}| \le |N_t| + |\tDelta_{e,\dett,t}| \le S_0+2S_a$.
Next we bound $|\Delta_{\dett,t}|$.
Using the above discussion, $\Delta_t \subseteq \Sset_{t-1}(2) \cup \Aset_t \setminus \Rset_t$. Using (\ref{sseteq_2}) for $j=2$, the RHS equals $\Sset_t(2) \cup \Iset_t(2) \setminus \Dset_t(1)$.
Apply Lemma \ref{detectcond_modcs} with $S_N = S_0$, $S_{\Delta_e}=S_a$, $S_\Delta = 2S_a$, and with $b_1 = 2r$ (so that $\Delta_1 \subseteq \Iset_t(2)$).
Since conditions \ref{measmodel} and \ref{add_del} of the theorem hold, all the undetected elements of $\Iset_{t}(2)$ will definitely get detected at time $t$. Thus $\Delta_{\dett,t} \subseteq \Sset_t(2) \setminus \Dset_{t}(1)$.
Since $\Dset_{t}(1) \subseteq \Sset_{t}(2)$, so $|\Delta_{\dett,t}| \le |\Sset_t(2)| - |\Dset_{t}(1)|= S_a$.

Consider the deletion step. Apply  Lemma \ref{truedelscond} with $S_T = S_0+2S_a$, $S_\Delta = S_a$.
Since condition \ref{measmod_delta} holds, $\delta_{S_0+2S_a} < 1/2$. Since $\Delta_{\dett,t} \subseteq \Sset_t(2) \setminus \Dset_{t}(1)$, so $\|x_{\Delta_{\dett}}\| \le \sqrt{S_a}r$. Since condition \ref{delthresh} also holds, all elements of $\tDelta_{e,\dett,t}$ will get deleted. Thus $|\tDelta_{e,t}|=0$. Thus $|\tT_t| \le |N_t| + |\tDelta_{e,t}| \le S_0$.
%Use $\Delta_{\dett,t} \subseteq \Sset_t(2) \setminus \Dset_{t}(1) \subseteq \Sset_t(2)$ to bound $\|x_{\Delta_{\dett}}\|$ by $\sqrt{2S_a}r$.
%
Next, we bound $|\tDelta_{t}|$.
Apply Lemma \ref{nofalsedelscond} with $S_T = S_0+2S_a$, $S_\Delta = S_a$, $b_1 = 2r$. %Use $\Delta_{\dett,t} \subseteq \Sset_t(2) \setminus \Dset_{t}(1) \subseteq \Sset_t(2)$ to bound $\|x_{\Delta_{\dett}}\|$ by $\sqrt{2S_a}r$.
Since $\Delta_{\dett,t} \subseteq \Sset_t(2) \setminus \Dset_{t}(1)$, so $\|x_{\Delta_{\dett}}\| \le \sqrt{S_a}r$.
By Lemma \ref{nofalsedelscond}, to ensure that all elements of $T_{\dett,t}$ with magnitude greater than or equal to $b_1=2r$ do not get falsely deleted, we need $\delta_{S_0+2S_a} < 1/2$ and $2r > \alpha_{del} + \sqrt{2} \eps + 2 \theta_{S_0+2S_a,S_a} \sqrt{S_a} r$. From condition \ref{delthresh}, $\alpha_{del} =  \sqrt{2} \eps + 2 \theta_{S_0+2S_a,S_a} \sqrt{S_a} r$. Thus, we need $\delta_{S_0+2S_a} < 1/2$ and $2r > 2(\sqrt{2} \eps + 2 \theta_{S_0+2S_a,S_a} \sqrt{S_a} r)$. $\delta_{S_0+2S_a} < 1/2$ holds since condition \ref{measmod_delta} holds. The second condition holds since condition \ref{measmod_theta} and condition \ref{add_del} ($r \ge G_2$) hold. Thus, we can ensure that all elements of $T_{\dett,t}$ with magnitude greater than or equal to $b_1=2r$ do not get falsely deleted.
But nothing can be said about the elements smaller than $2r$. In the worst case $\tDelta_t$ may contain all of these elements, i.e. it may be equal to $\Sset_t(2)$. Thus, $\tDelta_t \subseteq \Sset_t(2)$ and so $|\tDelta_t| \le 2S_a$.%since all these conditions hold,

 This finishes the proof of the first claim. To prove the second and third claims for any $t>0$: use the first claim  for $t-1$ and the arguments from the paragraphs above to show that the second and third claim hold for $t$.
The fourth claim follows directly from the first claim and fact \ref{errls1} of Proposition \ref{prop1} (applied with  $x \equiv \xhat_t$, $T \equiv \tT_t$, $\Delta \equiv \tDelta_t$). The fifth claim follows directly from the second claim and Corollary \ref{modcs_cs_bnd}.
%The last two claims follow easily using the first two claims and the mod-cs error bound or LS step error bound respectively.

%The first claim of the corollary follows using fact \ref{errls1} of Sec. \ref{stab_modcs}, the bounds on $\tT$ and $\tDelta$ and using $\|x_{\tDelta}\| \le \sm(2)$. The second follows using Theorem \ref{modcsbnd}, $|N|=S_0$ and the bounds on $|\Delta_e|$ and $|\Delta|$.

%and $\Delta \subseteq \Sset_{t-1}(2) \cup \Aset_t \setminus \Rset_t$.

\subsection{Appendix: Generalized version of Theorem \ref{stabres_modcs}}% and Corollary \ref{stabres_modcs_cor}
\label{stabres_modcs_gen}

\begin{theorem}[Stability of modified-CS with add-LS-del] % (generalized)
Assume Signal Model \ref{sigmod2} and  $\|w_t\| \le \eps$. Let $e_t:=(x_t - \xhat_{\dett,t})_{T_{\dett,t}}$. Assume that $\|e_t\|_\infty \le \|e_t\| / \sqrt{S_a}$ at all $t$ (the LS step error is spread out enough). If for some $1 \le d_0 \le d$, %the following hold%
\ben
\item {\em (addition and deletion thresholds) }
\ben
\item $\alpha_{add}$ is large enough so that there are at most $f$ false additions per unit time,
\label{addthresh}

%\item $\alpha_{del}  = \sqrt{2} \eps +  2  k_3  \sqrt{S_a} \theta_{S_0+S_a+f,k_2} r  $,
\item $\alpha_{del}  = \sqrt{\frac{2}{S_a}} \eps +  2  k_3  \theta_{S_0+S_a+f,k_2} r  $,
\label{delthresh}
\een

%\item {\em (no. of measurements, $n$) } $n$ is large enough so that
\item {\em (support size, support change rate)} $S_0,S_a$ satisfy
\ben
\item $\delta_{S_0 + S_a(1 + k_1) } < (\sqrt{2}-1)/2$ ,
\item $\delta_{S_0+S_a + f} < 1/2$,%$k^a S_a \le \sinf$, $S_0+S_a + f \le \sone$,
%\item $\theta_{S_0+S_a+f,k_2} < \frac{1}{2} \frac{d_0}{4k_3\sqrt{S_a} }$,
\item $\theta_{S_0+S_a+f,k_2} < \frac{1}{2} \frac{d_0}{4k_3}$,
\label{theta_ass_0}
\een
\label{measmodel}

\item {\em (new element increase rate) } $r \ge \max(G_1,G_2)$, where
\label{add_del}
\bea
G_1 \sdefn \frac{ \alpha_{add} + 8.79\eps }{d_0}  \nn \\ %C_1(S_0 + S_a + k_1) \eps
%G_2 \sdefn \frac{2\sqrt{2} \eps}{d_0 -  4k_3 \sqrt{S_a} \theta_{S_0+S_a+f,k_2} }  \ \ \ \ \ \ \
G_2 \sdefn \frac{2\sqrt{2} \eps}{\sqrt{S_a} (d_0 -  4k_3 \theta_{S_0+S_a+f,k_2}) }  \ \ \ \ \ \ \
\eea
\item {\em (initial time)} at $t=0$, $n_0$ is large enough to ensure that $\tDelta  \subseteq \Sset_0(d_0)$, $|\tDelta| \le (2d_0-2)S_a$,  $|\tDelta_e| =0$, $|\tT| \le S_0$,%$|\tDelta| \le (2d_0-2)S_a$, we use enough measurements,
\label{initass}
\een
where
\bea
k_1 \sdefn \max(1,2d_0-2) \nn \\
k_2 \sdefn \max(0,2d_0-3) \nn \\
k_3 \sdefn \sqrt{ \sum_{j=1}^{d_0-1} j^2 +  \sum_{j=1}^{d_0-2} j^2 }  %\frac{S_a (\sum_{j=1}^{d_0-1} j^2 + \sum_{j=1}^{d_0-2} j^2) M^2}{d^2}  \|(x)_{\Sset(d_0)}\|
\eea
then, at all $t \ge 0$,
\ben
\item  $|\tT| \le S_0$, $|\tDelta_e| =0$, and $\tDelta \subseteq \Sset_t(d_0)$ and so $|\tDelta| \le (2d_0-2)S_a$,

\item $|T| \le S_0$, $|\Delta_e| \le S_a$, and $|\Delta| \le k_1$,

\item $|T_\dett| \le S_0+S_a+f$, $|\Delta_{e,\dett}| \le S_a+f$, and $|\tDelta_\dett| \le k_2$

\item $\|x_t-\xhat_t\| \le \sqrt{2} \eps +  k_3  \sqrt{S_a} (2\theta_{S_0,(2d_0-2)S_a}+1) r  $

\item $\|x_t - \xhat_{t,modcs}\| \le C_1(S_0+S_a + k_1) \eps \le 8.79 \eps$.
\een
\end{theorem}

 %Notice that  $\|x_\Delta\|  \le \sqrt{|\Delta_1|} b + \|x_{\Delta_2}\|$.

\subsection{Proof of Lemma \ref{detectcond_lscs}}
\label{proof_detectcond_lscs}
From Lemma \ref{lscs_bnd}, if $\|w\| \le \eps$, $\delta_{2|\Delta|} < \sqrt{2}-1$ and $\delta_{|T|} < 1/2$, then $\|x - \xhat_{CSres}\| \le C'(|T|,|\Delta|) \eps +  \theta_{|T|,|\Delta|} C''(|T|,|\Delta|) \|x_\Delta\|$.
Using the fact that $\|x_\Delta\| \le \sqrt{|\Delta_1|} b + \|x_{\Delta_2}\|$; fact \ref{det1} of Proposition \ref{prop1}; and the fact that for all $i \in \Delta_1$, $|x_i| \ge \gamma b$, we can conclude that all $i \in \Delta_1$ will get detected if $\delta_{2|\Delta|} < (\sqrt{2}-1)/2$, $\delta_{|T|} < 1/2$ and $\alpha_{add} + C' \eps + \theta C'' \|x_{\Delta_2}\| + \theta C'' \sqrt{|\Delta_1|} b < \gamma b$. Using $\|x_{\Delta_2}\| \le \kappa b$ and $|\Delta_1| \le S_{\Delta_1}$, this last inequality holds if $\theta C'' \le {\gamma}/{2(\sqrt{S_{\Delta_1}} + \kappa)}$ and $\frac{ \alpha_{add} + C'\eps }{\gamma - \theta C'' (\sqrt{S_{\Delta_1}} + \kappa)} < b$.
%${\theta} \sqrt{|\Delta_1|}C'' < {\gamma}/2$ and $\frac{\alpha_{add} + C' \eps + \theta  C'' \|x_{\Delta_2}\|}{\gamma - \theta \sqrt{|\Delta_1|} C''} < b$.%
%
Since we only know that $|T| \le S_T$, $|\Delta| \le S_\Delta$,  $|\Delta_1| \le S_{\Delta_1}$ and $\|x_{\Delta_2}\| \le \kappa$, we need the above four inequalities to hold for all values of $|T|,|\Delta|,|\Delta_1|,\|x_{\Delta_2}\|$ satisfying these upper bounds. This leads to the conclusion of the lemma. Notice that the LHS's of all the required inequalities, except the last one, are non-decreasing functions of  $|\Delta|,|T|,|\Delta_1|$ and thus the lemma just uses their upper bounds. The LHS of the last one is non-decreasing in $|T|,|\Delta_1|,\|x_{\Delta_2}\|$, but is not monotonic in $|\Delta|$ (since $C'(|T|,|\Delta|)$ is not monotonic in $|\Delta|$). Hence we explicitly maximize over $|\Delta| \le S_\Delta$.
%This is not true for $|\Delta|$, since $C''(|T|,|\Delta|)$ is not monotonic in $|\Delta|$. Hence we need to explicitly maximize over $|\Delta| \le S_\Delta$. ,\|x_{\Delta_2}\|
$\blacksquare$

%\input{lscs}
%\input{algos} \input{stab_modcs}

%old: %\input{mainfile} %\input{extrastuff}

%\appendix
%\vspace{-0.1in}
%\subsection{Generative Model for Signal Model \ref{sigmod2}}
%\label{generativemodel}
%\vspace{-0.05in}
%See long version \cite{long}.
%
%\vspace{-0.15in}
%\subsection{Proof of Theorem \ref{stabres_simple_modcs}}% and Corollary \ref{stabres_modcs_cor}
%\label{proof_simple_modcs}
%\vspace{-0.05in}
%See long version \cite{long}.
%
%\vspace{-0.15in}
%\subsection{Proof of Theorem \ref{stabres_modcs}}% and Corollary \ref{stabres_modcs_cor}
%\label{proof_addLSdel_modcs}
%\vspace{-0.05in}
%See long version \cite{long}.
%
%\vspace{-0.15in}
%\subsection{Generalized version of Theorem \ref{stabres_modcs}}% and Corollary \ref{stabres_modcs_cor}
%\label{stabres_modcs_gen}
%\vspace{-0.05in}
%See long version \cite{long}.
%\vspace{-0.2in}

\bibliographystyle{IEEEbib}
\bibliography{tipnewpfmt_kfcsfullpap}

\end{document}